\documentclass[pdflatex,sn-mathphys-num]{sn-jnl}

\usepackage{graphicx}
\usepackage{amsmath,amssymb,amsfonts}
\usepackage{amsthm}
\usepackage{mathtools}
\usepackage{bm}
\usepackage{comment}

\theoremstyle{thmstyleone}
\newtheorem{theorem}{Theorem}[section]
\newtheorem{lemma}[theorem]{Lemma}
\newtheorem{proposition}[theorem]{Proposition}
\newtheorem{corollary}[theorem]{Corollary}
\theoremstyle{thmstyletwo}

\theoremstyle{thmstylethree}

\newcommand{\R}{\mathbb{R}}
\newcommand{\cV}{\mathcal{V}}
\newcommand{\cK}{\mathcal{K}}
\DeclareMathOperator{\Int}{Int}
\DeclareMathOperator{\conv}{conv}
\newcommand{\norm}[1]{\lVert #1\rVert}

\begin{document}

\title[Union of Translates of a Convex Body]{Union of Translates of a Convex Body}

\author*[1]{\fnm{Alec} \sur{Shelley}}\email{ams01@stanford.edu}
\author[2]{\fnm{Daniel} \sur{Tartakovsky}}

\affil*[1]{\orgdiv{Department of Applied Physics}, \orgname{Stanford University}, \orgaddress{\city{Stanford}, \state{CA}, \country{USA}}}
\affil[2]{\orgdiv{Department of Energy Science and Engineering}, \orgname{Stanford University}, \orgaddress{\city{Stanford}, \state{CA}, \country{USA}}}

\abstract{\unboldmath Let $K\subset\mathbb{R}^d$ be a strictly convex body and let $Z\subset\mathbb{R}^d$ be a finite set of sites in general position. We consider the problem of calculating the volume $\mu_d$ of $\bigcup_{z\in Z}(z-K)$. Let $V_z$ be the Voronoi cell of $z\in Z$ with respect to the convex distance function induced by $K$, and let $N$ be the nerve of the Voronoi cells. When $K$ is a ball, a classical result reduces the inclusion-exclusion formula to a sum over $N$:
\[\mu_d\Bigl(\bigcup_{z\in Z}(z-K)\Bigr)=\sum_{J\in N}(-1)^{|J|+1}\,\mu_d\Bigl(\bigcap_{z\in J}(z-K)\Bigr).\]
When may the hypothesis that $K$ is a ball be dropped? We show that for $d\leq 2$ the reduced formula holds for every strictly convex $K$, while for each $d\geq 3$ there is a strictly convex $K$ for which it fails. For $x\in \mathbb{R}^d$, let $\chi_x$ be the Euler characteristic of the subcomplex of $N$ induced by the sites lying in $K+x$. A sufficient condition for the reduced formula to hold is $\chi_x \equiv 1$ on the union. When $d=3$ and $K$ is $C^4$, we show that $\chi_x\equiv 1$ on the union for all $Z$ in general position if and only if $K$ is an ellipsoid. The proof relies on the Fubini--Pick affine invariant.}



\maketitle
%\tableofcontents
\section{Introduction}

\begin{figure}[t]
\centering
\begin{minipage}{0.485\textwidth}\centering
\includegraphics[width=\textwidth]{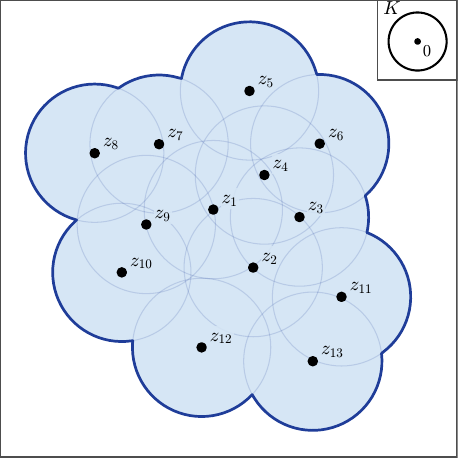}\\[2pt]
{\small (a) disk-shaped $K$}
\end{minipage}\hfill
\begin{minipage}{0.485\textwidth}\centering
\includegraphics[width=\textwidth]{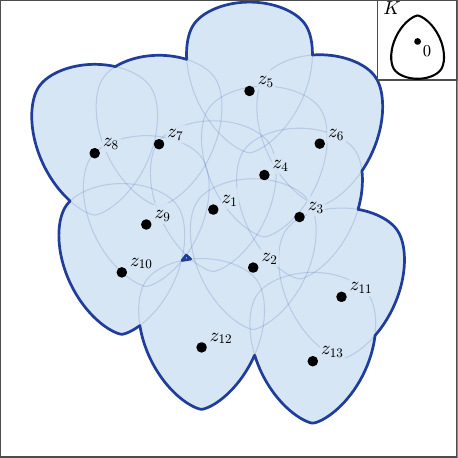}\\[2pt]
{\small (b) egg-shaped $K$}
\end{minipage}\\[4pt]
\includegraphics[width=0.95\textwidth]{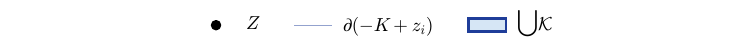}
\caption{Unions of translates of a convex body $-K \subset \mathbb{R}^2$, taken to be disk-shaped in (a) and egg-shaped in (b). The centers $Z = \{z_1,\dots,z_{13}\}$ are labeled with black dots and are shared between (a) and (b). We wish to calculate the area of the union, shown in light blue.}
\label{fig:union}
\end{figure}

We study volume of the union of translates of a strictly convex body $K\subset \mathbb{R}^d$. Let $Z =\{z_i\}_{i=1}^m \subset \mathbb{R}^d$, and $\mathcal{K} \coloneqq \{z_i-K\}_{i=1}^m$. We wish to calculate the volume $\mu_d \left(\bigcup \mathcal{K} \right)$, as illustrated in Figure \ref{fig:union}.

The inclusion-exclusion formula expresses this measure as an alternating sum over the $2^m-1$ nonempty subsets of $\mathcal K$, which is prohibitively slow for large $m$. When $K$ is a ball and $Z$ is in general position, a classical result reduces the sum to just those terms corresponding to the nerve of the Voronoi cells \cite{naiman1992inclusion} (see also \cite{edelsbrunner1995union, attali2007inclusion}). Each remaining term is then an intersection of at most $d+1$ balls, and the number of terms is $\mathcal O(m^{\lceil d/2\rceil})$ \cite[Theorem 4.4]{aurenhammer2000voronoi}: an exponential speedup. Since the reduced formula is stated entirely in terms of the Voronoi diagram, it is natural to ask whether the ball hypothesis on $K$ may be dropped. 

Voronoi diagrams for convex distance functions, defined through the Minkowski functional of a strictly convex body \cite{ma2000bisectors, chew1985voronoi, corbalan1996geometry, aurenhammer2013voronoi}, are a well-studied generalization of Euclidean Voronoi diagrams. One might conjecture that, if $K$ is a strictly convex body, the topology of the corresponding Voronoi diagram is sufficiently close to the Euclidean case for the reduced formula to hold. We show that this is the case precisely when $d \leq 2$.

In the plane, part of the answer is already known. Every pseudodisk collection admits a two-dimensional abstract simplicial complex, embeddable in the plane, over which the inclusion-exclusion sum may be taken~\cite{edelsbrunner1997inclusion}. Translates of $-K$ in general position form a pseudodisk collection, so this applies to our problem, but the complex of \cite{edelsbrunner1997inclusion} is generally not unique and need not be the Voronoi nerve. We show that the nerve of planar Voronoi cells serves the same purpose.

We now fix notation. For any $a\in \mathbb{R}$ and $b\in \mathbb{R}^d$, let $aK+b \coloneqq \{ax+b:x \in K\}$. Assume that $0 \in \Int K$, and let $p_K:\mathbb{R}^d\rightarrow \mathbb{R}^+$ be the Minkowski functional of $K$ given by 
\begin{equation}
p_K(x) \coloneqq \min \{r\geq0: x \in rK\}.
\end{equation}
The value $p_K(y-x)$ is known as a \emph{convex distance function} from $x$ to $y$ \cite{chew1985voronoi}, and we use it to define Voronoi cells. Let the Voronoi cell of $z_i$ be 
\begin{equation}
V_i = V(z_i) \coloneqq \{x \in \mathbb{R}^d : p_K(z_i-x) \leq p_K(z_j-x) \text{ for all }j \}.
\end{equation}
Let $\mathcal{V} \coloneqq  \{V(z_i)\}_{i=1}^m$ be the family of Voronoi cells. 

We say that $Z$ is in \emph{general position} when every intersection of $d+2$ of its Voronoi cells is empty. Dimension counting shows that this is true except on a nowhere dense set.

\begin{figure}[htbp]
\centering
\begin{minipage}{0.485\textwidth}\centering
\includegraphics[width=\textwidth]{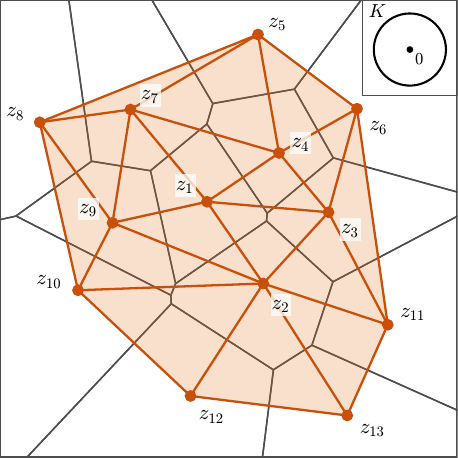}\\[2pt]
{\small (a) disk-shaped $K$}
\end{minipage}\hfill
\begin{minipage}{0.485\textwidth}\centering
\includegraphics[width=\textwidth]{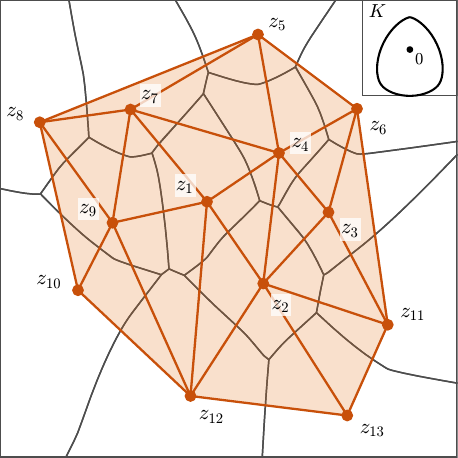}\\[2pt]
{\small (b) egg-shaped $K$}
\end{minipage}\\[4pt]
\includegraphics[width=0.95\textwidth]{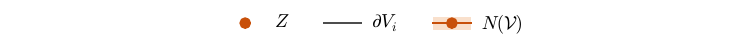}
\caption{The Voronoi cells $V_i=V(z_i)$ of a set $Z=\{z_1,\dots,z_{13}\}$ of sites in the plane, for two bodies $K$ (insets, with the origin marked). Each $z_i$ is in a unique Voronoi cell whose borders are drawn in black. The nerve $N(\mathcal{V})$ is drawn on the sites: a segment $z_iz_j$ when $V_i\cap V_j\neq\emptyset$, and a shaded triangle $z_iz_jz_k$ when $V_i\cap V_j\cap V_k\neq\emptyset$. The terms in (\ref{maineq}) correspond to the vertices, edges, and faces of $N(\mathcal{V})$, totaling $13+29+17=59$ terms for both (a) and (b). By contrast, the ordinary inclusion-exclusion formula would sum over $2^{13}-1=8191$ terms. Note that in (a) the Voronoi cells are convex, while in (b) they are only star convex.}
\label{fig:nerve}
\end{figure}

For any family of sets $\mathcal{A}$, let $\bigcap \mathcal{A} \coloneqq \bigcap_{A \in \mathcal{A}} A$. Define the \emph{nerve} $N(\mathcal{A})$ of $\mathcal{A}$ to record its non-empty intersections:
\begin{equation}
N(\mathcal{A}) \coloneqq \{X \subseteq \mathcal{A}: \bigcap X \neq \emptyset \}.
\end{equation}
For each element $J\in N(\mathcal{V})$, let the corresponding bodies be $\mathcal{K}_J = \{z_i-K: V(z_i) \in J\}$. Finally, let $|J|$ be the cardinality of a finite set $J$.
Figure \ref{fig:nerve} shows the Voronoi cells and the nerve $N(\mathcal{V})$ for two choices of $K$.

The \emph{inclusion-exclusion formula} reads
\begin{equation}
\mu_d\left(\bigcup \mathcal{K}\right) = \sum_{J \subseteq \mathcal{K}}(-1)^{|J|+1}\mu_d \left( \bigcap J \right).
\end{equation}
For any $d$ and when $K$ is a ball, the following simplification holds \cite{naiman1992inclusion}:
\begin{equation}
\mu_d\left(\bigcup \mathcal{K}\right) = \sum_{J \in N(\mathcal{V})}(-1)^{|J|+1}\mu_d \left( \bigcap \mathcal{K}_J \right).
\label{maineq}
\end{equation}
Our first main result is that the ball hypothesis on $K$ may be dropped precisely when $d \leq 2$:

\begin{theorem}
\label{thm1}
Formula (\ref{maineq}) holds for all strictly convex $K$ and $Z$ in general position if and only if $d \leq 2$.
\end{theorem}

\begin{figure}[htbp]
\centering
\begin{minipage}{0.485\textwidth}\centering
\includegraphics[width=\textwidth]{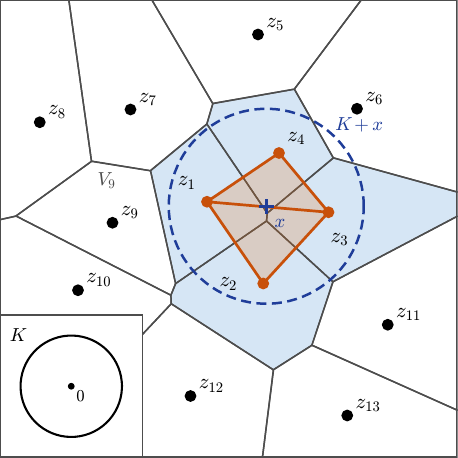}\\[2pt]
{\small (a) disk-shaped $K$}
\end{minipage}\hfill
\begin{minipage}{0.485\textwidth}\centering
\includegraphics[width=\textwidth]{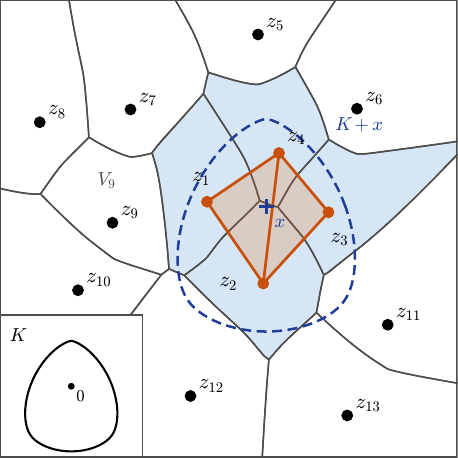}\\[2pt]
{\small (b) egg-shaped $K$}
\end{minipage}\\[4pt]
\includegraphics[width=0.95\textwidth]{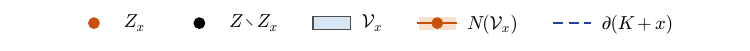}
\caption{The sites and the bodies of Figure \ref{fig:nerve}, with a point $x$. The dashed curve bounds the translate $K+x$. The sites in $K+x$ form $Z_x=\{z_1,z_2,z_3,z_4\}$. Their Voronoi cells form $\mathcal{V}_x$, and $N(\mathcal{V}_x)$ is the subcomplex of $N(\mathcal{V})$ induced by these cells. In both (a) and (b), $\chi_x=4-5+2=1$.}
\label{fig:defs}
\end{figure}

Since $Z$ is in general position, $J \in N(\mathcal{V})$ has $|J| \leq d+1$, so each term in (\ref{maineq}) involves only $d+1$-wise intersections. In the $d=2$ case, this lets us calculate the area of the union using only the area of the individual bodies and of their pairwise and triple intersections. Furthermore, when $d=2$ and $m\geq 3$, Corollary \ref{cor:size} shows that the number of terms in (\ref{maineq}) is no greater than $6m-11$. 

Following \cite[Section 2]{naiman1992inclusion} (see also \cite{attali2007inclusion}), we derive a sufficient condition for (\ref{maineq}). For any $x\in \mathbb{R}^d$, let $Z_x = \{z_i : z_i \in K+x\}$ and $\mathcal{V}_x \coloneqq \{V(z_i) : z_i \in Z_x\}$. Then $N(\mathcal{V}_x)$ is the subcomplex of $N(\mathcal{V})$ induced by $\mathcal{V}_x$. Consider its Euler characteristic $\chi_x$, defined by $\chi_x \coloneqq \sum_{J \in N(\mathcal{V}_x)} (-1)^{|J|+1}$. See Figure \ref{fig:defs} for a visual.
We show that (\ref{maineq}) holds whenever $\chi_x = \mathbf{1}_{\bigcup \mathcal{K}}(x)$. We begin by writing
\begin{equation}
\begin{split}
\mu_d\left(\bigcup \mathcal{K}\right) = \int_{\mathbb{R}^d} \mathbf{1}_{\bigcup \mathcal{K}}(x) dx = \int_{\mathbb{R}^d} \chi_x dx
= \int_{\mathbb{R}^d} \sum_{J \in N(\mathcal{V}_x)} (-1)^{|J|+1} dx.
\end{split}
\end{equation}
Since $z_i \in K+x$ if and only if $x \in z_i-K$, for any $x\in \mathbb{R}^d$, $J \in N(\mathcal{V}_x)$ if and only if $J \in N(\mathcal{V})$ and $x \in \bigcap \mathcal{K}_J$. This gives
\begin{equation}
 \mu_d\left(\bigcup \mathcal{K}\right)= \sum_{J \in N(\mathcal{V})}(-1)^{|J|+1} \int_{\mathbb{R}^d}\mathbf{1}_{\bigcap \mathcal{K}_J}(x)dx,
\end{equation}
which is (\ref{maineq}). Existing approaches for the Euclidean case \cite[Lemma 3.1]{naiman1992inclusion}, \cite{attali2007inclusion} prove that $\chi_x = \mathbf{1}_{\cup \mathcal{K}}(x)$ by showing that $\bigcup \mathcal{V}_x$ is contractible whenever $x \in\bigcup \mathcal{K}$, and is empty otherwise. This uses the nerve theorem from algebraic topology \cite{borsuk1948imbedding}, relying on the $V_i$ and their intersections being contractible. For our proof of Theorem \ref{thm1} when $d=2$, we circumvent the nerve theorem and instead rely on planarity. When $d \geq 3$ and $K$ is non-ellipsoidal, non-planar intersections of homothets \cite{goodey1981intersections,Goodey_1983} lead to the possibility of four Voronoi cells intersecting at multiple points \cite{santos1996delaunay, icking1999bisectors,icking1995convex}, so the hypotheses of the nerve theorem fail. The $d\geq 3$ case of Theorem \ref{thm1} shows that this is not merely a barrier to a proof: We construct an explicit non-ellipsoidal $K\subset\mathbb{R}^d$ and $Z$ for which (\ref{maineq}) fails.

Theorem \ref{thm1} in $d\geq 3$ says that (\ref{maineq}) fails for \emph{some} non-ellipsoidal $K$ and $Z$, but it does not say what happens for \emph{every} non-ellipsoidal $K$. Our second main result helps answer this question in $d=3$. We prove that when $K \subset \mathbb{R}^3$ is $C^4$, the equation $\chi_x = \mathbf{1}_{\cup \mathcal{K}}(x)$ characterizes ellipsoids:

\begin{theorem}
\label{thm2}
Let $d=3$ and $K$ be a $C^4$, strictly convex body. Then $\chi_x = \mathbf{1}_{\bigcup \mathcal{K}}(x)$ holds for all $x \in \mathbb{R}^3$ and $Z$ in general position if and only if $K$ is an ellipsoid.
\end{theorem}

The if direction follows from an affine transformation of the case when $K$ is a ball \cite{naiman1992inclusion}. For the only if direction, we use the Fubini-Pick cubic form, an affine invariant in the $3$-jet of a surface which vanishes identically if and only if the surface is a quadric. Since the only compact quadrics are ellipsoids, the form vanishes identically if and only if $K$ is an ellipsoid. The non-vanishing of the Fubini--Pick cubic form \cite{NomizuSasaki1994} at some point on $\partial K$ will let us explicitly construct $Z$ so that $\chi_x =2$ for all $x$ in an open subset of $\bigcup \mathcal{K}$, proving Theorem \ref{thm2}. 

Motivated by Theorem \ref{thm2}, we conjecture that when $d\geq3$ and $K$ is not an ellipsoid, (\ref{maineq}) fails for an open set of configurations $Z$. Otherwise, the errors from the regions where $\chi_x \neq 1$ would have to cancel exactly, even after perturbing $Z$. 

 In Section \ref{sec_prelim} we review properties of Voronoi cells induced by a Minkowski functional $p_K$, and in Section \ref{sec_2D} we focus on such properties unique to the plane. We prove Theorem \ref{thm1} for the $d\leq 2$ case in Section \ref{sec_thm1} and the $d\geq 3$ case in Section \ref{sec:example}, then prove Theorem \ref{thm2} in Section \ref{sec:3d}.

\section{Preliminaries}
\label{sec_prelim}

In this section, we show that some familiar properties of Voronoi cells hold when $p_K$ is used in place the Euclidean distance. These properties hold in any dimension. Several of these are results from \cite{martini2001geometry, martini2004geometry}.

\begin{lemma}
\label{lemmabasic}
    The function $p_K$ has positive homogeneity, satisfies the strict triangle inequality, and is strictly convex. That is, for any $x,y \in \mathbb{R}^d$, $\lambda\geq 0$, and $t \in (0,1)$,
\begin{equation}
\label{positive homogenity}
p_K(\lambda x) = \lambda p_K(x),
\end{equation}
\begin{equation}
\label{triangle ineqality}
p_K(x+y) \leq p_K(x)+p_K(y),
\end{equation}
\begin{equation}
\label{strict convexity}
p_K((1-t)x+ty) \leq (1-t)p_K(x)+tp_K(y),
\end{equation}
Where in the last two equations, equality holds if and only if there is some $a\geq 0$ so that $x=ay$.
\end{lemma}
\begin{proof} (\ref{positive homogenity}) is clear from the definition of $p_K$, and with it (\ref{triangle ineqality}) and (\ref{strict convexity}) are equivalent. Since $K$ is strictly convex, $(1-t)x+ty \in \text{Int}(K)$, from which (\ref{strict convexity}) follows.
\end{proof}
Let $\lVert \cdot \rVert$ be the Euclidean norm in $\mathbb{R}^d$, and $d(\cdot, \cdot)$ the Euclidean distance.

\begin{lemma}
\label{lemmaLipschitz}
There are constants $c,C>0$ so that for any distinct $x,y \in \mathbb{R}^d$, we have $c\lVert x\rVert\leq p_K(x) \leq C\lVert x\rVert$ and $\lvert p_K(x)-p_K(y)\rvert \leq C\lVert x-y\rVert $.
\end{lemma}
\begin{proof}
From the triangle inequality, we arrive at the reverse triangle inequality, 
\begin{equation}
\label{reversetriangleinequality}
    \lvert p_K(x)-p_K(y)\rvert \leq \max \{p_K(x-y),p_K(y-x)\}.
\end{equation}
Let $r:S^{d-1} \rightarrow \mathbb{R}^+$ be the radial function of $K$, given by $r(\omega) = \max\{\lambda\geq 0:\lambda\omega \in K\}$. Since $K$ is a compact body with the origin in its interior, there are constants $c,C$ so that $1/C\leq r(\omega)\leq 1/c$. By definition, $r(\omega) = 1/p_K(\omega)$. Combining this with positive homogeneity proves the first claim. After upper bounding the RHS of (\ref{reversetriangleinequality}) by $C\lVert x-y\rVert $, the second claim follows.
\end{proof}

In the Euclidean case, the Voronoi cells $V(z)$ are convex polytopes. Otherwise, we have the following weaker result:
\begin{lemma}
Each $V_i$ is strictly star-shaped. That is, for $x \in V_i$ we have $(x,z_i] \subset \Int V_i$.
\label{star}
\end{lemma}
\begin{proof}
Let $r_t \coloneq (1-t)x+tz_i$ and  $K_t \coloneq p_K(z_i-r_t)K+r_t$. Using positive homogeneity, we get $K_t=(1-t)p_K(z_i-x)K+r_t = (1-t)(p_K(z_i-x)K+x-z_i)+z_i$. After translation, $K_t$ is the minimizing body in the definition of $p_K(z_i-r_t)$, so $z_i\in \partial K_t$. By strict convexity of $K$, for any $0\leq t_1 < t_2 \leq1$, we have $K_{t_2}\setminus \{z_i\} \subset \text{Int}(K_{t_1})$. In particular, $K_t\setminus\{z_i\} \subset \text{Int}(K_0)$. If $z_j \neq z_i$, then $z_j \notin \text{Int}(K_0)$, so $z_j \notin K_t\setminus\{z_i\}$. By the definition of $p_K(z_i-r_t)$ and $V_i$, we conclude that $(1-t)x+tz_i \in \text{Int}(V_i)$.
\end{proof}

For distinct $x,y \in \mathbb{R}^d$, let $L(x,y) \coloneqq \{x' \in \mathbb{R}^d: p_K(x-x')=p_K(y-x')\}$. 

\begin{lemma}
For distinct $x, y \in \mathbb{R}^d$, there is a hyperplane $H \subset \mathbb{R}^d$ with unit normal vector $\nu$, and a continuous function $g: H \rightarrow \mathbb{R} $ so that $L(x,y) = \{h +g(h)\nu : h\in H\}$. In addition, this makes $L(x,y)$ homeomorphic to $\mathbb{R}^{d-1}$.
\label{planehom}
\end{lemma}
The proof is given in Appendix \ref{app:2d}.

The following lemma demonstrates how the Voronoi cells cover $\mathbb{R}^d$, and it is the reason we required $K$ to be strictly convex instead of just convex.
\begin{lemma} Each $\partial V_i$ has empty interior. For $i\neq j$, $\Int V_i \cap V_j = \emptyset$.
\end{lemma}

\begin{proof}
The first part follows immediately from Lemma \ref{planehom}. The second follows Lemma \ref{star} and the fact that $x \in \Int V_i$ if and only if $p_K(z_i-x) < p_K(z_j-x)$ for all $z_j \neq z_i$.
\end{proof}

\begin{proposition}\label{prop:conn}
$\bigcup\cV_x$ is connected.  Consequently, the graph on $Z_x$ in which $z_i\sim z_j$ when $V_{ij}\ne\emptyset$ is connected.
\end{proposition}

\begin{proof}
The site nearest to $x$ lies in $Z_x$, so $x\in\bigcup\cV_x$, and every connected component of $\bigcup\cV_x$ is a union of Voronoi cells. Take distinct components $C_1\ni x$ and $C_2$,  and take $z_{i_2}\in Z_x$ with $V_{i_2}\subseteq C_2$. Let $r(t)\coloneq(1-t)x+tz_{i_2}$ and $t_0:=\sup\{t\in[0,1]:\ r(t)\in C_1\}$. Since  $C_1$ is closed, $r(t_0)\in C_1$ and $t_0<1$.  Choose a sequence $t_n\downarrow t_0$ with $t_n>t_0$ and all $r(t_n)$ in one cell $V_{i_3}$.  Then $V_{i_3}$ contains $r(t_0)\in C_1$ and the points $r(t_n)\notin C_1$, so $V_{i_3}\not\subseteq C_1$. Thus $V_{i_3} \notin \mathcal{V}_x$, giving 
\[
 p_K(z_{i_3}-x)>1\ge p_K(z_{i_2}-x).
\]
Therefore the argument of Lemma \ref{star} applied to $p_K(z_{i_2}-x)K+x$ gives $p_K(z_{i_2}-r(t))<p_K(z_{i_3}-r(t))$ for all $t\in[0,1]$. This contradicts $r(t_n)\in V_{i_3}$, proving the first claim.

For the second claim, suppose $Z_x=Y_1\sqcup Y_2$ with both parts nonempty and $V_{ij}=\varnothing$ whenever $z_i\in Y_1$ and $z_j\in Y_2$.  Then $\bigcup_{z_i\in Y_1}V_i$ and $\bigcup_{z_j\in Y_2}V_j$ are nonempty disjoint closed sets covering $\bigcup\cV_x$, contradicting connectedness.
\end{proof}

\section{Two-Dimensional Preliminaries} In this section we exhibit nice topological properties of planar Euclidean Voronoi cells which are shared by their non-Euclidean counterparts. Several analogous results from this section in $\mathbb{R}^3$ are false, see for example \cite{icking1999bisectors}.
\label{sec_2D}

We call a transformation of a set $A \subset \mathbb{R}^d$ a \emph{homothety} if it is a translation or it is of the form $A \rightarrow \lambda (A-v)+v$ for some $\lambda >0$ and $v\in \mathbb{R}^d$. In the second case, we call $v$ the center. We call the image of $A$ under any homothety a \emph{homothet} of A.
\begin{lemma}
\label{homothety}
Let $K' \subset \mathbb{R}^2$ be compact and strictly convex, and let $x,y,z$ be distinct points in $\partial K'$. For any homothety $h\neq I$, $h(\{x,y,z\})$ can't all lie in $\partial K'$. 
\end{lemma}
The proof is given in Appendix \ref{app:2d}.

\begin{corollary}
When $d=2$ and for any homothety $h\neq I$, $\partial K \cap h(\partial K)$ contains at most two points.
\label{sphereintcor}
\end{corollary}
\begin{proof}
Suppose by contradiction that 
\(p_1,p_2,p_3\in \partial K\cap h(\partial K)\) are distinct.
Set \(q_\ell=h^{-1}(p_\ell)\). Then \(q_1,q_2,q_3\in \partial K\), and
\(h(q_1),h(q_2),h(q_3)\in \partial K\), which is impossible by Lemma \ref{homothety}. 
\end{proof}

We denote intersections of the Voronoi cells by $V_{i_1...i_n} = V_{i_1} \cap ... \cap V_{i_n}$. This notation will always imply that $i_1,...,i_n$ are distinct.

\begin{proposition}
Let $d=2$. Then $V_{ij}$ is connected, $V_{ijk}$ contains at most one point, and when $Z$ is in general position, $V_{ijkl} = \emptyset$.
\label{connectedprop}
\end{proposition}
\begin{proof}
If $|Z| = 2$, then Lemma \ref{planehom} shows that $V_{ij}$ is homeomorphic to $\mathbb{R}$ and thus connected. If $|Z|>2$, assume $V_{ijk}$ contains two points $u_1\neq u_2$. Then there exists $\lambda_1, \lambda_2 >0$ so that $\{z_i,z_j,z_k\} \subset \partial (\lambda_1K+u_1)\cap \partial (\lambda_2K+u_2)$, contradicting Corollary \ref{sphereintcor}. Thus $V_{ijk}$ is either empty or a single point, which we call $u$. We have proven the second claim, and the third claim follows from the definition.

We now prove the first claim for $|Z|>2$. Let $g:\mathbb{R} \rightarrow L(z_i,z_j)$ be the homeomorphism from Lemma \ref{planehom}, and let $f_m:\mathbb{R}^2 \rightarrow \mathbb{R}$ be given by $f_m(w) \coloneqq p_K(z_i-w)-p_K(z_m-w)$. The function $f_m(g(t))$ is continuous, and by the previous part of the proposition, has at most one zero, call it $t_m$. Thus $(f_m\circ g)^{-1}(-\infty,0]$ is connected. Let $A_{m} \coloneqq \{w\in L(z_i,z_j) : \max\{p_K(z_i-w), p_K(z_j-w)\}\leq p_K(z_m-w)\}$. We have just shown that $g^{-1}(A_m)$ is connected. Therefore $g^{-1}(V_{ij}) = \bigcap_m A_m$ is also connected, and since $g$ is continuous, $V_{ij}$ is connected. We have proven the first claim, and we are done.
\end{proof}

It is clear from the above lemma that $Z$ is in general position except on a nowhere dense set.

For $u\in\R^d$ let $s(u):=\min_{i}p_K(z_i-u)$ and $S(u):=\{i:\ p_K(z_i-u)=s(u)\}$. By positive homogeneity, $u\in V_i$ if and only if $i\in S(u)$.
We call $u$ a \emph{vertex} of $\mathcal{V}$ if $|S(u)|=3$. We write $u_{ijk}$ for a vertex with $S(u) = \{i,j,k\}$. 

\begin{lemma}\label{lem:arc}
Let $d=2$, $Z$ be in general position, and $V_{ij}\ne\varnothing$. Then $V_{ij}$ is homeomorphic to a nondegenerate closed segment, a ray, or a line, and the endpoints of $V_{ij}$ are vertices of $\mathcal{V}$.
\end{lemma}

\begin{proof}
Let $g:\R\to L(z_i,z_j)$ be the homeomorphism of Lemma \ref{planehom}. By Proposition \ref{connectedprop}, $V_{ij}$ is closed and connected, so $g^{-1}(V_{ij})$ is a closed interval. For the first claim, we must rule out a single point.  Assume that $V_{ij} = \{u\}$.

If $S(u)=\{i,j\}$, then $p_K(z_i-u)=p_K(z_j-u)<p_K(z_l-u)$ for every $l\notin\{i,j\}$. By continuity, all points of $L(z_i,z_j)$ close enough to $u$ also lie in $V_{ij}$, a contradiction.

If $S(u) = \{i,j,k\}$, then there is some small enough neighborhood $D$ around $u$ such that $V_i$, $V_j$ and $V_k$ are the only Voronoi cells intersecting $D$. Otherwise, by closedness, we would get $|S(u)| > 3$. Parametrize space in polar coordinates $(r,\theta)$ centered at $z_k$. Let $u = (r_0,\theta_0)$, and consider the box $\Gamma \subset D$ with vertical sides $[(r_0-\delta_r,\theta_0 \pm\delta_\theta),(r_0+\delta_r, \theta_0 \pm \delta_\theta )]$ and horizontal sides $[(r_0\pm \delta_r, \theta_0-\delta_\theta), (r_0 \pm \delta_r, \theta_0 + \delta_\theta)]$. By strict star-convexity and for $\delta_r$ small enough, $(r_0-\delta_r ,\theta_0) \in \Int V_k$ and $(r_0+\delta_r ,\theta_0) \notin V_k$. Since $p_K$ is continuous, there is a $\delta_\theta$ small enough that the lower horizontal side lies entirely within $\Int V_k$, while the upper horizontal side lies entirely outside $V_k$. Consider $(r_0-\delta_r+r,\theta_0-\delta_\theta) \in \Gamma$. By star-convexity, there is a largest $r$ such that this point lies in $V_k$, call it $r_l$. Let $w_l = (r_0-\delta_r+r_l,\theta_0-\delta_\theta)$. This is the left endpoint of $V_k \cap \Gamma$. Let the right endpoint be $w_r$, defined similarly. Let $\Gamma^+$ be the open upper arc in $\Gamma$ from $w_l$ to $w_r$, and let $\Gamma^-$ be the closed lower arc, so that $V_k \cap \Gamma = \Gamma^-$. The segments $[z_i,u) \in V_i$ and $[z_j,u) \in V_j$ intersect $\Gamma$, and this intersection must lie in $\Gamma^+$. Since $V_i$ and $V_j$ form a closed cover of $\Gamma^+$, they must intersect in $\Gamma^+$, contradicting $V_{ij} = \{u\}$.

If $|S(u)|>3$, then this contradicts Proposition \ref{connectedprop}. Thus the first claim is proved. For the second claim, let $u$ be an endpoint of $V_{ij}$. Take a sequence of points $u_n \rightarrow u$ with $u_n \in L(z_i,z_j)$ but $u_n \notin V_{ij}$. Since $|\mathcal{V}| < \infty$, some subsequence of the $u_n$ must lie in a common Voronoi cell, call it $V_k$. Since $V_k$ is closed, we conclude $u$ is a vertex of $\mathcal{V}$.
\end{proof}

\begin{lemma}\label{lem:ray}
Let $W\subseteq Z$ and take $z \in Z$ with $z\notin\conv(W)$.  Then there is a ray from $z$ which lies outside $\bigcup_{w \in W}V(w)$.
\end{lemma}

\begin{proof}
Since $z \notin \conv(W)$, we may pick a unit vector $\nu$ with $\langle\nu,w\rangle > \langle\nu,z\rangle$ for all $w\in W$.  Let $\kappa\in\partial K$ maximize $\langle\nu,\cdot\rangle$ over $K$, and let $r(t):=z-t\kappa$. Then $p_K(z-r(t))=t$, while
\[
 t K+r(t)=z+t(K-\kappa)\subseteq\{\langle\nu,\cdot\rangle\le\langle\nu,z\rangle\}.
\]
Thus each $w\in W$ lies outside $t K+r(t)$, so $p_K(z-r(t)) < p_K(w-r(t)).$
\end{proof}

\section{Proof of Theorem \ref{thm1} for \texorpdfstring{$d \leq 2$}{d <= 2}}
\label{sec_thm1}
When $d=1$, every convex body is a translate of a ball, so by the affine invariance of $N(\mathcal V_x)$ noted at the start of Section \ref{sec:3d}, (\ref{maineq}) follows from the case of balls \cite{naiman1992inclusion}. The rest of the section proves the theorem for $d=2$.

Recall the definitions $Z_x\coloneq \{z_i:\ z_i\in K+x\}$, $\cV_x\coloneq \{V_i:\ z_i\in Z_x\}$, and that $\chi_x$ is the Euler characteristic of $N(\mathcal{V}_x)$. We fix $x \in \bigcup\mathcal{K}$, and construct a planar graph $\Gamma_x$ which corresponds to the dual graph of $\mathcal{V}_x$, modified by bending each edge at a point interior to the edge. We will show that the Euler characteristic of $\Gamma_x$ is $\chi_x$, and then we separately show that it is equal to $1$.

We identify each $J\in N(\cV_x)$ with the set of indices of its cells, so that $\{i,j\}\in N(\cV_x)$ means $z_i,z_j\in Z_x$ and $V_{ij}\ne\varnothing$. Write $N_k:=\{J\in N(\cV_x):\ |J|=k\}$.  By Proposition~\ref{connectedprop}, $N_k=\varnothing$ for $k\ge4$, so
\[
 \chi_x=|N_1|-|N_2|+|N_3|,
\]
and each $\{i,j,k\}\in N_3$ has exactly one vertex which we call $u_{ijk}$.

\begin{figure}[t]
\centering
\includegraphics[width=0.72\textwidth]{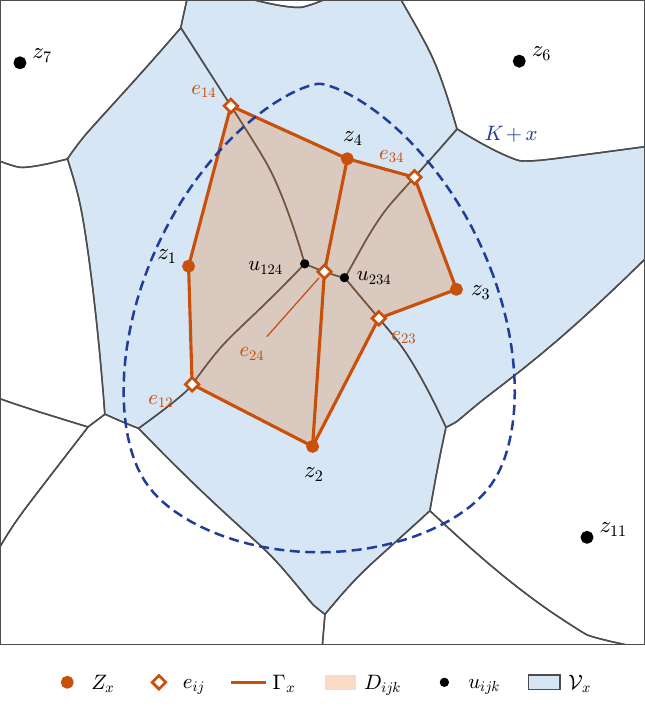}
\caption{The graph $\Gamma_x$ for the egg-shaped $K$, the sites and the point $x$ of Figure \ref{fig:defs}(b). For each intersecting pair of cells in $\cV_x$, a point $e_{ij}$ is chosen on the interior of $V_{ij}$, and $z_i$ and $z_j$ are joined by the bent path $E_{ij}=[z_i,e_{ij}]\cup[e_{ij},z_j]$. The interior of both segments are contained in the interior of one Voronoi cell, and $E_{ij}$ intersects $V_{ij}$ only at $e_{ij}$ (Proposition \ref{prop:graph}). Each bounded face $D_{ijk}$ of $\Gamma_x$ contains exactly one vertex of $\mathcal{V}$, $u_{ijk}$ (Proposition \ref{prop:faces}). Here $\Gamma_x$ has $4+5=9$ vertices, $2\cdot5=10$ edges, and $2$ bounded faces, so its Euler characteristic is $9-10+2=1$. This equals $\chi_x=4-5+2$.}
\label{fig:gamma}
\end{figure}

If $K$ were a disc, then each $V_i$ would be convex, so we could construct $\Gamma_x$ by letting its edges be the segments of the form $[z_i,z_j]$ for $\{i,j\} \in N_2$. However, when $K$ is not a disc, the $V_i$ are merely star-convex, so $[z_i,z_j]$ might exit $V_i \cap V_j$. We therefore modify this construction in the following way: For each $\{i,j\}\in N_2$, choose a non $\mathcal{V}$-vertex $e_{ij}\in V_{ij}$ (possible by Lemma~\ref{lem:arc}), set $E_{ij}:=[z_i,e_{ij}]\cup[e_{ij},z_j]$, and let
\[
 \Gamma_x\coloneq \bigcup_{\{i,j\}\in N_2}E_{ij}.
\]
Figure \ref{fig:gamma} shows an example.

We define the \emph{faces} of $\Gamma_x$ to be the connected components of $\R^2\setminus\Gamma_x$. Since $\Gamma_x$ is bounded, it has exactly one unbounded face. Also note that a connected subset of $\R^2 \setminus \Gamma_x$ intersecting a face $F$ is contained in $F$.

The following proposition shows that each vertex of $\Gamma_x$ of the form $z_i$ corresponds to a face $V_i$ of $\mathcal{V}_x$, and each bent arc $E_{ij}$ of $\Gamma_x$ corresponds to an edge $V_{ij}$ of $\mathcal{V}_x$. We will later show that faces of $\Gamma_x$ correspond to vertices of $\mathcal{V}_x$. In this way, one should think of $\Gamma_x$ as a kind of dual graph to $\mathcal{V}_x$.

\begin{proposition}\label{prop:graph}
$\Gamma_x$ is a connected plane graph with vertices $Z_x\cup\{e_{ij}\}$ and edges $\{[z_i,e_{ij}], [e_{ij},z_j]\}$. Moreover
\begin{enumerate}[(b)]
\item[(a)] $\Gamma_x \cap V_i = \bigcup_j [z_i,e_{ij}]$.
\item[(b)] $\Gamma_x \cap V_{ij} = \{e_{ij}\}$.
\item[(c)] If $z_i \notin Z_x$, then $\Gamma_x \cap V_i = \emptyset$.
\end{enumerate}
The right hand sides of $(a),(b)$ are interpreted as the empty set if no such $e_{ij}$ exists.
\end{proposition}

\begin{proof}
$\Gamma_x$ is connected as a consequence of Proposition \ref{prop:conn}. If $z_i \in Z_x$, then $z_i \in \Gamma_x$ if any $e_{ij}$ exists. If no ${e_{ij}}$ exists, then by second claim of Proposition \ref{prop:conn}, $Z_x = \{z_i\}$. This shows that $Z_x \subseteq \Gamma_x$, so $\Gamma_x$ has vertex set $Z_x \cup \{e_{ij}\}$. By construction, $\Gamma_x$ has the claimed edges.

By strict star-convexity, $[z_i,e_{ij}) \subset \Int V_i$, so the definition of $\Gamma_x$ and $e_{ij}$ gives (a) and (b). By construction, $E_{ij} \cap V_k = \emptyset$ for $k \notin \{i,j\}$, so $z_i \in \Gamma_x$ only if $z_i \in Z_x$. By (a), if $\Gamma_x \cap V_i \neq \emptyset$, then $z_i \in \Gamma_x$, and from the previous sentence $z_i \in Z_x$. This proves (c). Next we prove that $\Gamma_x$ is a planar graph.

Since $[z_i,e_{J}) \subset \Int V_i$, two edges $[z_i,e_J]$ and $[z_{i'},e_{J'}]$ with $i\ne i'$ meet at most at their common endpoint $e_J=e_{J'}$, a vertex of $\Gamma_x$. Two different edges $[z_i,e_{ij}]$, $[z_i,e_{ik}]$ with $i\ne j$ meet only at $z_i$ unless $e_{ij}$ and $e_{ik}$ lie on a common ray from $z_i$. In that case, assume WLOG $[z_i,e_{ij}] \subset [z_i,e_{ik}]$. By star-convexity, $[z_i,e_{ik}) \subset \Int V_i$, contradicting $e_{ij} \in V_{ij}$. We have shown the edges of $\Gamma_x$ intersect only at the vertices of $\Gamma_x$. Finally, we note the vertices of $\Gamma_x$ are exactly the endpoints of its edges. We conclude that $\Gamma_x$ is a planar graph.
\end{proof}

The following lemma is the only time when membership of $Z_x$ is used in the proof of Theorem \ref{thm1}.
\begin{lemma}\label{lem:noout}
If $z_i\notin Z_x$, then $V_i$ is contained in the unbounded face of $\Gamma_x$.
\end{lemma}

\begin{proof}
By Proposition \ref{prop:graph} (c), $V_i$ is disjoint from $\Gamma_x$, so $V_i$ lies in a single face $F$. Since $K$ is convex, $z_i\notin Z_x$ implies $z_i\notin\conv Z_x$. The ray of Lemma~\ref{lem:ray} with $W=Z_x$ and starting at $z_i$ is disjoint from $\bigcup \mathcal{V}_x \supset \Gamma_x$. Since the ray is contained in $F$, $F$ must be unbounded.
\end{proof}

\begin{lemma}\label{lem:vertex}
Let $\Lambda$ be a set of indices with $z_l\in Z_x$ for each $l\in \Lambda$, and let $F$ be a bounded face of $\Gamma_x$ with $F\cap V_l=\varnothing$ for every $l\notin \Lambda$.  Then $F$ contains a vertex of $\mathcal{V}$, $u$, with $S(u)\subseteq \Lambda$.
\end{lemma}

\begin{proof}
$F$ is open, nonempty, and contained in $\bigcup_{l\in \Lambda}V_l$, so it contains a point $q$ in the interior of $V_l$ for some $l \in \Lambda$, and lying off of the rays from $z_l$ through the points $e_{lj}$. Let ${P}$ be the open ray from $z_l$ to $q$, excluding $z_l$. By Proposition \ref{prop:graph} (a), ${P} \cap V_l$ is disjoint from $\Gamma_x$. Since ${P} \cap V_l$ is a connected set intersecting $F$, it is contained in $F$, and thus bounded. Let $p \coloneq {P} \cap \partial V_l \in V_{kl}$. By the assumption on $F$, $k \in \Lambda$. Let $C$ be component of $V_{kl} \setminus e_{kl}$ containing $p$. By Proposition \ref{prop:graph} (b), $C \subset F$, so $C$ is bounded. By Lemma \ref{lem:arc}, we conclude that the endpoint of $C$, $u$, is a vertex of $\mathcal{V}$. If we let $S(u) = \{j,k,l\}$, then by the hypothesis on $F$ we have $j \in \Lambda$.
\end{proof}

The next proposition shows that faces of $\Gamma_x$ correspond to vertices of $\mathcal{V}_x$.

\begin{proposition}\label{prop:faces}
Assume that $\{i,j,k\}\in N_3$. Then the closed curve $H_{ijk}:=E_{ij}\cup E_{jk}\cup E_{ki}$ is a Jordan curve whose inside $D_{ijk}$ is a bounded face of $\Gamma_x$ containing a vertex $u_{ijk}$ of $\mathcal{V}$. Every bounded face of $\Gamma_x$ arises in this way, and $\{i,j,k\}\mapsto D_{ijk}$ is a bijection from $N_3$ onto the set of bounded faces of $\Gamma_x$.
\end{proposition}

\begin{proof}
Let us write $H \coloneq H_{ijk}$ and $D \coloneq D_{ijk}$. By Proposition \ref{prop:graph}, the arcs $E_{ij},E_{jk}$ and $E_{ki}$ meet pairwise only in the shared endpoints $z_i,z_j,z_k$, so $H$ is a Jordan curve. By the Jordan curve theorem, we can take $D$ to be its inside. $H$ consists of six half-open segments lying in $\Int V_i$, $\Int V_j$, $\Int V_k$, together with the three points $e_{ij},e_{jk},e_{ki}$. Since interiors of distinct cells are disjoint and $S(e_{ij})=\{i,j\}$, we have $H\cap V_l=\varnothing$ for every $l\notin\{i,j,k\}$.

We show that $D \cap Z = \emptyset$. The sites $z_i,z_j,z_k$ lie on $H$.  Let $l\notin\{i,j,k\}$, and assume that $z_l\in\conv\{z_i,z_j,z_k\}$. Then by strict convexity of $K$, $z_l$ lies in the interior of $s(u_{ijk})K+u$. Thus $p_K(z_l-u_{ijk})<s(u_{ijk})$, contradicting the definition of $s(u_{ijk})$. Thus $z_l\notin\conv\{z_i,z_j,z_k\}$, so the ray of Lemma~\ref{lem:ray} using $W=\{z_i,z_j,z_k\}$ and starting at $z_l$ avoids $V_i\cup V_j\cup V_k\supseteq H$. Thus it is contained in $D$, which is a contradiction since $D$ is bounded. 

$V_l$ is connected, we have shown that it is disjoint from $H$, and it contains $z_l\notin\overline D = H \cup D$. Thus we have shown
\begin{equation}
V_l \cap \overline{D} = \emptyset \quad \text{for each }l \notin \{i,j,k\}.
\label{eqn_excl}
\end{equation}

Next we prove that $D$ is a face of $\Gamma_x$. Since $D$ is open and connected with $\partial D=H\subseteq\Gamma_x$, it suffices to check $D\cap\Gamma_x=\varnothing$. We have already shown $z_l \notin D$. Any $e_J$ with $J\not\subseteq\{i,j,k\}$ lies in $V_l$ for some $l\notin\{i,j,k\}$, as does any edge $[z_l,e_J]$ with $l\notin\{i,j,k\}$. Both are excluded by (\ref{eqn_excl}). The remaining edges are of the form $[z_i,e_{il}]$ with $l\notin\{j,k\}$. Since $\Gamma_x$ is a planar graph, the segment $[z_i,e_{il}]$ lies in $\Int V_i\cup\{e_{il}\}$, hence misses the four segments of $H$ contained in $V_j\cup V_k$, and it meets $[z_i,e_{ij}]$ and $[z_i,e_{ik}]$ only at $z_i$. Thus $(z_i,e_{il}]$ is connected and disjoint from $H$, and it contains $e_{il}\in V_l$ which lies outside $\overline{D}$ by (\ref{eqn_excl}). Hence $(z_i,e_{il}]$ lies outside $\overline{D}$ as well. We have shown that $D$ is a face of $\Gamma_x$.
 
Next we show $u_{ijk} \in D$. By (\ref{eqn_excl}) we may apply Lemma~\ref{lem:vertex} to $D$ with $\Lambda =\{i,j,k\}$ to produce a vertex $u'\in D$ with $S(u')\subseteq\{i,j,k\}$. Thus $S(u')=\{i,j,k\}$ and $u'=u_{ijk}$ is the desired vertex of $\Gamma_x$.

Next we prove bijectivity of the map $\{i,j,k\} \mapsto D_{ijk}$.  Injectivity is immediate from $D_{ijk} \cap Z = \{z_i,z_j,z_k\}$. By Lemmas~\ref{lem:noout} and~\ref{lem:vertex}, applied with $W$ taken as the set of indices of $Z_x$, a bounded face $F$ contains a vertex $u_{ijk}$ with $\{i,j,k\}\in N_3$. Then $F$ and $D_{ijk}$ are faces sharing a point, hence equal.
\end{proof}

\begin{proposition}\label{prop:chi}
$\chi_x=1$ for every $x\in\bigcup\cK$.
\end{proposition}

\begin{proof}
Euler's formula says that for connected plane graphs with $n$ vertices, $m$ edges, and $f$ faces (including the unbounded one), $n-m+f=2$. For $\Gamma_x$, we have $n=|N_1|+|N_2|$, $m=2|N_2|$, and by Proposition~\ref{prop:faces}, $f=1+|N_3|$. Euler's formula then gives $|N_1|-|N_2|+|N_3|=1$.
\end{proof}

\begin{proof}[Proof of Theorem \ref{thm1} for $d=2$:]
If $x\notin\bigcup\cK$ then $\cV_x=\varnothing$ and $\chi_x=0$. Otherwise $\chi_x=1$ by Proposition~\ref{prop:chi}.  Hence $\chi_x=\mathbf 1_{\bigcup\cK}(x)$ for every $x$, and following the argument in the introduction,
\[
 \mu_2\Bigl(\bigcup\cK\Bigr)=\int_{\R^2}\chi_x\,dx=\sum_{J\in N(\cV)}(-1)^{|J|+1}\mu_2\Bigl(\bigcap\cK_J\Bigr).
\]

\end{proof}

\begin{corollary}
\label{cor:size}
Let $d=2$, $m\ge3$, and $Z$ be in general position. Then $N(\mathcal V)$ has $m$ vertices, at most $3m-6$ edges, and at most $2m-5$ triangles. Consequently the sum in (\ref{maineq}) has at most $6m-11$ terms.
\end{corollary}
\begin{proof}
The construction of this section applies verbatim with $Z$ in place of $Z_x$: $\bigcup\mathcal V=\R^2$ is connected, so Proposition~\ref{prop:conn} holds, and Lemma~\ref{lem:noout} is vacuous. Suppressing the degree-two vertices $e_{ij}$, Proposition~\ref{prop:graph} exhibits the $1$-skeleton of $N(\mathcal V)$ as a simple plane graph on the $m$ vertices $Z$, and Proposition~\ref{prop:faces} identifies the triangles of $N(\mathcal V)$ with its bounded faces. A simple plane graph on $m\ge3$ vertices has at most $3m-6$ edges and at most $2m-5$ bounded faces.
\end{proof}

\section{Proof of Theorem \ref{thm1} for \texorpdfstring{$d\geq 3$}{d >= 3}}
\label{sec:example}

Theorem \ref{thm2} shows that for a non-ellipsoidal $K$ the identity $\chi_x=\mathbf 1_{\bigcup\cK}(x)$ fails, but it does not by itself show that (\ref{maineq}) fails. In this section we give an explicit $K$ and $Z$ for which (\ref{maineq}) fails.

Write points of $\R^3$ as $(u,v,w)$ and let $r\coloneqq\sqrt{u^2+v^2}$. For $\varepsilon>0$ consider the smoothed bicone given by
\begin{equation}
\label{eq:Keps}
K_\varepsilon\coloneqq\bigl\{(u,v,w)\in\R^3:\ \sqrt{r^2+\varepsilon^2}+\sqrt{w^2+\varepsilon^2}\le1+\varepsilon\bigr\}.
\end{equation}
 $K_\varepsilon$ is a smooth, symmetric, strictly convex body, and it is clearly not an ellipsoid. As $\varepsilon\to0$, $K_\varepsilon$ converges to the bicone $K_0\coloneqq\{r+|w|\le1\}$. More precisely,
\begin{equation}
\label{eq:sandwich}
\{r+|w|\le1-\varepsilon\}\subset K_\varepsilon\subset\{r+|w|\le1+\varepsilon\},
\quad\text{so}\quad
\frac{p_{K_0}}{1+\varepsilon}\le p_{K_\varepsilon}\le\frac{p_{K_0}}{1-\varepsilon}.
\end{equation}
Let $s\coloneqq19/20$ and $R\coloneqq s/\sqrt3\approx0.5485$, and take
\begin{equation}
\label{eq:sites}
z_1=(0,0,0),\quad z_{2,3}=\Bigl(\mp\tfrac s2,\ \tfrac{\sqrt3}{2}s,\ 0\Bigr),\quad z_{4,5}=\Bigl(0,\ R,\ \pm\tfrac12\Bigr),\quad x\coloneqq z_1.
\end{equation}
Thus $z_1,z_2,z_3$ form an equilateral triangle of side $s$ in the plane $\{w=0\}$ with circumcenter $g\coloneqq(0,R,0)$ and circumradius $R$, and $z_4,z_5$ lie on the vertical line $\ell$ through $g$. We will use the inequalities $s<1$ and $1/2<R$, and assume that $\varepsilon < R-1/2$.

By the defining formula for $K_{\varepsilon}$, we may write $p_{K_{\varepsilon}}(u,v,w)=F(r,|w|)$ for a function $F$ which is strictly increasing in each argument and has $F(a,b) = F(b,a)$.

\emph{(i) $Z_x=\{z_1,z_2,z_3\}$.} By (\ref{eq:sandwich}), $p_{K_{\varepsilon}}(z_2-x)=p_{K_{\varepsilon}}(z_3-x)\le s/(1-\varepsilon)<1$ and $p_{K_{\varepsilon}}(z_4-x)=p_{K_{\varepsilon}}(z_5-x)\ge (R+\tfrac12)/(1+\varepsilon)>1$. By continuity, $Z_y=\{z_1,z_2,z_3\}$ for all $y$ in some open neighborhood of $x$.

\emph{(ii) $V_{123}=\emptyset$.} For $y=(q,w)$ with $q\in\R^2$ and $i\le3$, we have $p_{K_{\varepsilon}}(z_i-y)=F(\lVert q-z_i\rVert,|w|)$, so $y\in V_{123}$ only if $y\in\ell$. If $y=(g,w)\in\ell$, then for $i \leq 3$ we have $p_{K_{\varepsilon}}(z_i-y)=F(R,|w|)$, while for one of $j=4,5$ we have $p_{K_{\varepsilon}}(z_j-y)=F(0,t)$ with $t=\bigl||w|-\tfrac12\bigr|<\max\{R,|w|\}$. Hence $y\notin V_{123}$.

\emph{(iii) $N(\cV)=\{J\ne\emptyset:\ J\not\supseteq\{1,2,3\}\}$.} The inclusion $\subseteq$ is (ii). Conversely, a nerve is closed under taking subsets, and every $J$ on the right lies in one of $\{1,2,4,5\}$, $\{1,3,4,5\}$, $\{2,3,4,5\}$. By symmetry, it suffices to show that $V_{1245}\ne\emptyset$. Let $m\coloneqq\tfrac12(z_1+z_2)$. Every $y$ on the segment $[m,g)$ lies in the plane $\{w=0\}$ and on the perpendicular bisector of $z_1z_2$, so $p_{K_{\varepsilon}}(z_4-y)=p_{K_{\varepsilon}}(z_5-y)$ and $p_{K_{\varepsilon}}(z_1-y)=p_{K_{\varepsilon}}(z_2-y)<p_{K_{\varepsilon}}(z_3-y)$. The continuous function $\varphi(y)\coloneqq p_{K_{\varepsilon}}(z_4-y)-p_{K_{\varepsilon}}(z_1-y)$ satisfies
\[
\varphi(m)=F(\tfrac R2,\tfrac12)-F(\tfrac s2,0)>F(\tfrac12,0)-F(\tfrac s2,0)>0,\qquad \varphi(g)=F(\tfrac12,0)-F(R,0)<0.
\]
Thus $\varphi$ vanishes at some $y\in(m,g)$, and this $y$ lies in $V_{1245}$.

In particular $V_{12345}=\emptyset$, so $Z$ is in general position. For $y\in\bigcup\cK$, (iii) gives
\[
\chi_y=\sum_{\emptyset\ne J\subseteq Z_y}(-1)^{|J|+1}-\sum_{\{1,2,3\}\subseteq J\subseteq Z_y}(-1)^{|J|+1}=1-\mathbf 1_{Z_y = \{z_1,z_2,z_3\}}(y).
\]
Here the first sum is $1$. The second sum vanishes unless $\{1,2,3\}\subseteq Z_y$, in which case it equals $\sum_{T\subseteq Z_y\cap\{4,5\}}(-1)^{|T|}$, which is $1$ if $Z_y\cap\{4,5\}=\emptyset$ and $0$ otherwise. By the computation following Theorem \ref{thm1}, the right-hand side of (\ref{maineq}) equals $\int_{\R^3}\chi_y\,dy$. Hence
\begin{equation}
\begin{split}
\mu_3\Bigl(\bigcup\cK\Bigr)-\sum_{J\in N(\cV)}(-1)^{|J|+1}\mu_3\Bigl(\bigcap\cK_J\Bigr) &=\int_{\R^3}\bigl(\mathbf 1_{\bigcup\cK}(y)-\chi_y\bigr)\,dy
\\ &=\mu_3\{y: Z_y=\{z_1,z_2,z_3\}\}>0,
\end{split}
\end{equation}
where in the last inequality we used (i).

The example extends to every $d\ge3$. Write points of $\R^d$ as $(a,t)$ with $a\in\R^3$ and $t\in\R^{d-3}$. Let $K'\coloneqq\{(a,t):\ p_{K_\varepsilon}(a)^2+\lVert t\rVert^2\le1\}$, a strictly convex body with $p_{K'}(a,t)=\bigl(p_{K_\varepsilon}(a)^2+\lVert t\rVert^2\bigr)^{1/2}$. Take the sites $Z=\{(z_i,0)\}_{i=1}^5$. For every $(a,t)$, the distances $p_{K'}\bigl((z_i,0)-(a,t)\bigr)$ are ordered in the same way as the $p_{K_\varepsilon}(z_i-a)$, so the Voronoi cells are $V_i\times\R^{d-3}$ and the nerve is unchanged from the $d=3$ case. The computation above then shows that the error in (\ref{maineq}) is $\mu_d\{y:\ Z_y=\{(z_1,0),(z_2,0),(z_3,0)\}\}$, which is positive because this set contains a neighborhood of $(x,0)$.

\section{Proof of Theorem \ref{thm2}}
\label{sec:3d}
\subsection{Overview}
If $J$ is an index set in $\mathcal{V}$, and $H$ is a homothet of $K$ with $z_i \in \partial H$ for each $i \in J$ and $Z \cap \Int H = \emptyset$, then we call $H$ a \emph{witness} of $J$. Note that $V_J\neq\emptyset$ if and only if $J$ has a witness. We use this to define $N(\mathcal{V}_x)$ even when $0 \notin \Int K$. Then for any invertible affine transformation $Av = Mv+b$, $N(\mathcal{V}_x)$ is preserved by the map $K \mapsto AK$, $Z \mapsto AZ$, $x \mapsto Mx$. 

Assume that $Z$ is in general position. By the previous paragraph and the fact that $\chi_x = \mathbf{1}_{\cup \mathcal{K}}$ when $K$ is a ball \cite{naiman1992inclusion}, we have $\chi_x = \mathbf{1}_{\cup \mathcal{K}}$ when $K$ is an ellipsoid. The rest of this section will prove the other direction of Theorem \ref{thm2}.

Let us assume that $d=3$ and $K$ is $C^4$ and non-ellipsoidal. We construct a point $x$ and sites 
\begin{equation}
Z=\{z_1,z_2,z_3,z_4,b_-,b_0,b_+\} \subset \mathbb{R}^3
\end{equation}
(so $m=7$, and $b_-,b_0,b_+$ stand for $z_5,z_6,z_7$) with the following properties: First, $Z_x = \{z_1,z_2,z_3,z_4\}$. Second, for each $1 \leq i<j<k \leq 4$, $V_{ijk}$ is non-empty. Third, $V_{1234} = \emptyset$. This gives
\begin{equation}
\chi_x = |N_1|-|N_2|+|N_3|-|N_4| = 4-6+4-0 = 2,
\end{equation}
proving Theorem \ref{thm2}. We accomplish this by showing that each triple in $\{1,2,3,4\}$ has a witness and that $\{1,2,3,4\}$ has none.

We use the Fubini--Pick cubic form \cite{blaschke1923affine}, a local affine invariant of $K$ lying in its $3$-jet. A version of the Maschke–Pick–Berwald theorem \cite{NomizuSasaki1994} says that the Fubini-Pick cubic form of $K$ vanishes on an open set exactly when $K$ is ellipsoidal there. To precisely state the theorem, we introduce some notation.

We write points of $\mathbb{R}^3$ as $(w,h)$ with $w=(w_1,w_2)\in\mathbb{R}^2$ and $h\in\mathbb{R}$, and call $h$ the \emph{height} of the point. 
Let $M$ be a $C^4$ compact surface in $\mathbb{R}^3$, and let $y \in M$ be a point with positive Gaussian curvature. After an affine transformation which moves $y$ to the origin, we may assume that $M$ is locally the graph of a $C^4$ function $f: \{w\in\mathbb{R}^2:\lVert w\rVert<r\} \rightarrow \mathbb{R}$, $r>0$, given by
\begin{equation}
\label{eqn_taylor1}
f(w) = \frac{1}{2}\lVert w \rVert^2 + \frac{1}{6}c(w) + \mathcal O(\lVert w\rVert^4).
\end{equation}
Let $\langle \alpha,w \rangle$ be the linear functional given by applying the Laplacian $\Delta$ to $c(w)/8$. 
There is a unique harmonic $c_0(w)$ so that
\begin{equation}
c(w) = c_0(w) + \lVert w \rVert ^2 \langle \alpha,w \rangle.
\end{equation}
Up to $\mathcal O(\lVert w \rVert^4)$, the shear $(w,h) \mapsto (w+\alpha h/3,h)$ gets rid of the $\langle \alpha, w \rangle$ term in $f$, leaving us with 
\begin{equation}
\begin{split}
f(w) &= \frac{1}{2}\lVert w\rVert ^2 +\frac{1}{6}c_0(w) + \mathcal O(\lVert w \rVert^4).
\label{eqn_normal_form}
\end{split}
\end{equation}
We call (\ref{eqn_normal_form}) the normal form of $M$ at $y$ (up to third order). Since $c_0$ is harmonic and cubic, it must be of the form $c_0(w) = {\kappa} \lVert w \rVert^3\cos(3 \phi- \phi_0).$ After rotating, this gives
\begin{equation}
\label{eqn_pick_form}
c_0(w) = \kappa P(w),\qquad P(w)\coloneqq\lVert w \rVert^3\cos(3 \phi) = w_1^3 - 3w_1w_2^2. 
\end{equation}
We call $c_0(w)$ the \emph{Fubini--Pick cubic form}. The only further affine transformations that fix the normal form are rotations, reflections, and rescalings $(w,h) \mapsto (\lambda w, \lambda^2 h)$, none of which changes whether $\kappa$ vanishes. Thus $\kappa = 0$ is invariant under affine transformations. This leads to the promised Maschke–Pick–Berwald  characterization of ellipsoids:

\begin{theorem}[Maschke–Pick–Berwald]\label{thm_pick_berwald}
Let $M\subset\mathbb{R}^3$ be a $C^4$, connected, compact surface. Then $M$ is an ellipsoid if and only if $\kappa=0$
at every point of $M$ with positive Gaussian curvature.
\end{theorem}
The theorem is classical, but we include an elementary proof in Appendix \ref{app:Pick}.

\subsection{Homothets}

By Theorem \ref{thm_pick_berwald}, $\partial K$ has a point of positive Gaussian curvature with $\kappa\neq0$. After an affine transformation, we may assume that this point is the origin, $K\subset\{h\ge0\}$, $\partial K$ has the normal form (\ref{eqn_normal_form}), (\ref{eqn_pick_form}) there, and, by the rescaling $(w,h) \mapsto (\kappa w,\kappa^2h)$, $\kappa=1$. So there is an $r_0>0$ such that for $\norm{w}\le r_0$ the lowest point of $K$ above $w$ is $(w,f(w))$, where
\begin{equation}
\label{eq:f}
f(w)=\tfrac12\norm{w}^2+\tfrac16P(w)+\mathcal O(\norm{w}^4)\qquad\text{and}\qquad D^2f(w)\ge\tfrac12\mathrm{Id}.
\end{equation}
Since $K$ is convex, has interior points, and meets the plane $\{h=0\}$ only at the origin, there are also $r_1\in(0,r_0]$ and $h_1>0$ with
\begin{equation}
\label{eq:inside}
\{(w,h):\ \norm{w}\le r_1,\ f(w)<h<h_1\}\subset\Int K.
\end{equation}

For $\varepsilon>0$, let
\[
\Phi_\varepsilon(w,h)\coloneqq(\varepsilon w,\varepsilon^2h),\qquad K_\varepsilon\coloneqq\Phi_\varepsilon^{-1}(K).
\]
Since $K_\varepsilon$ is an affine image of $K$, it suffices to construct $x$ and $Z$ for $K_\varepsilon$. There is an $\varepsilon_0>0$, depending only on $K$, such that for $0<\varepsilon < \varepsilon_0$, every statement in the rest of this section and in Appendix \ref{app:four} holds. The letter $C$ denotes a positive constant depending only on $K$, whose value may change from one occurrence to the next, and $X=\mathcal O(Y)$ means $\lVert X\rVert\le CY$.

Every homothet of $K_\varepsilon$ can be written uniquely as
\[
H=\sigma^{-1}K_\varepsilon+(\xi,a),\qquad\xi\in\R^2,\ a\in\R,\ \sigma>0.
\]
It is the image of $K$ under a map that multiplies horizontal distances by $1/(\sigma\varepsilon)$ and heights by $1/(\sigma\varepsilon^2)$. Hence, for $\norm{w-\xi}\le r_0/(\sigma\varepsilon)$, the lowest point of $H$ above $w$ is $(w,G(w))$, where
\begin{equation}
\label{eq:G}
\begin{gathered}
G(w)=a+\sigma^{-1}f_\varepsilon\bigl(\sigma(w-\xi)\bigr),\\
f_\varepsilon(y)\coloneqq\varepsilon^{-2}f(\varepsilon y)=\tfrac12\norm{y}^2+\tfrac{\varepsilon}{6}P(y)+\mathcal O(\varepsilon^2\norm{y}^4).
\end{gathered}
\end{equation}
We call $G$ the \emph{lower graph} of $H$. Expanding,
\begin{equation}
\label{eq:Gexp}
\begin{gathered}
G(w)=a+\tfrac{\sigma}{2}\norm{w-\xi}^2+\tfrac{\varepsilon\sigma^2}{6}P(w-\xi)+\mathcal O\bigl(\varepsilon^2\sigma^3\norm{w-\xi}^4\bigr),\\
D^2G(w)=\sigma\bigl(\mathrm{Id}+\mathcal O(\varepsilon\sigma\norm{w-\xi})\bigr).
\end{gathered}
\end{equation}
By (\ref{eq:inside}), we can decide whether a point lies in $H$ by comparing its height with $G$. For $\norm{w-\xi}\le r_1/(\sigma\varepsilon)$, we have
\begin{equation}
\label{eq:member}
(w,h)\notin H\ \text{ if }h<G(w),\qquad(w,h)\in\Int H\ \text{ if }G(w)<h<a+\frac{h_1}{\sigma\varepsilon^2}.
\end{equation}

\subsection{Construction}
\label{sec:config}
\begin{figure}[t]
\centering
\includegraphics[width=0.72\textwidth]{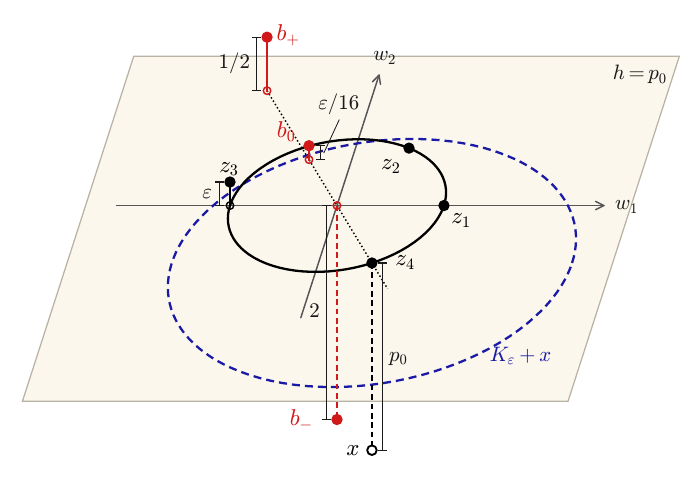}
\caption{The sites, seen from above and slightly tilted. The shaded plane is $h=p_0$, and each bracket gives the height of a site above or below it; the heights $\varepsilon$ and $\varepsilon/16$ are not drawn to scale. The sites $Z_x = \{z_1,\dots,z_4\}$ lie over the unit circle at $0^\circ,60^\circ,180^\circ,300^\circ$. The sites $b_0,b_+,b_-$, in red, lie over the dotted black line through $\omega_*$ and $\omega_4=-\omega_*$. The point $x=(\omega_4,0)$ is the lowest point of $K_\varepsilon+x$, and the dashed curve bounds the set of $w$ with $(w,p_0)\in K_\varepsilon+x$, up to $\mathcal O(\varepsilon)$.}
\label{fig:config}
\end{figure}

Let
\[
\omega_k\coloneqq(\cos\phi_k,\sin\phi_k)\quad\text{with}\quad(\phi_1,\phi_2,\phi_3,\phi_4)\coloneqq(0^\circ,60^\circ,180^\circ,300^\circ),
\]
let $\omega_*\coloneqq(\cos120^\circ,\sin120^\circ)$, and let $p_0\coloneqq\tfrac74$. Then
\[
\bigl(P(\omega_1),P(\omega_2),P(\omega_3),P(\omega_4)\bigr)=(1,-1,-1,-1),\qquad P(\omega_*)=1.
\]
Let $\gamma\coloneqq(-3,2,-1,2)$ and $\theta\coloneqq(0,0,1,0)$. Then $\sum_k\gamma_k=0$, $\sum_k\gamma_k\omega_k=0$, $\sum_k\gamma_kP(\omega_k)=-6$ and $\sum_k\gamma_k\theta_k=-1$. We use two facts about the points $\omega_k$.
\begin{itemize}[(E2)]
\item[(E1)] $\sum_k\gamma_k\ell(\omega_k)=0$ for every affine function $\ell$ on $\R^2$.
\item[(E2)] No three of the $\omega_k$ are collinear. Hence, for any three indices $k_1,k_2,k_3$ and affine $\ell$, we have  $\sup_{\norm{w}\le3}\lvert\ell(w)\rvert\le C\max_{1\leq i \leq 3}\lvert\ell(\omega_{k_i})\rvert$.
\end{itemize}
The sites are
\[
z_k\coloneqq(\omega_k,\,p_0+\varepsilon\theta_k)\quad(k=1,\dots,4),
\]
\[
b_0\coloneqq\bigl((1-\varepsilon)\omega_*,\,p_0+\tfrac1{16}\varepsilon\bigr),\qquad b_+\coloneqq\bigl(2\omega_*,\,p_0+\tfrac12\bigr),\qquad b_-\coloneqq\bigl(0,\,p_0-2\bigr).
\]
We let $x\coloneqq(\omega_4,0)$. See Figure \ref{fig:config} for an illustration of these definitions. The translate $K_\varepsilon+x$ is the homothet with $(\xi,a,\sigma)=(\omega_4,0,1)$, so by (\ref{eq:Gexp}) its lower graph is $\tfrac12\norm{w-\omega_4}^2+\mathcal O(\varepsilon)$ on $\norm{w}\le3$. At $\omega_1,\dots,\omega_4$, this equals $\tfrac12,\tfrac32,\tfrac32,0$ up to $\mathcal O(\varepsilon)$, all less than $p_0$. At $\omega_*$, $2\omega_*$ and $0$ it equals $2$, $\tfrac92$ and $\tfrac12$ up to $\mathcal O(\varepsilon)$, which exceed the heights of $b_0$, $b_+$ and $b_-$. By (\ref{eq:member}), $z_1,\dots,z_4\in\Int(K_\varepsilon+x)$ and $b_0,b_+,b_-\notin K_\varepsilon+x$. So $Z_x=\{z_1,\dots,z_4\}$.

\subsection{Admissible Homothets}

For every $\sigma>0$, the paraboloid $h=p_0+\tfrac{\sigma}{2}(\norm{w}^2-1)$ contains the unit circle at height $p_0$. We focus on homothets whose lower graph is close to such a paraboloid. Fix a constant $C_1$, depending only on $K$, that is at least as large as the constants $C$ in Lemmas \ref{lem:four} and \ref{lem:witness} below. We say that $H=\sigma^{-1}K_\varepsilon+(\xi,a)$ is \emph{admissible} if
\[
\tfrac14\le\sigma\le5,\qquad\norm{\xi}\le C_1\varepsilon,\qquad\bigl\lvert a-p_0+\tfrac{\sigma}{2}\bigr\rvert\le C_1\varepsilon.
\]
Let $H$ be admissible. For small enough $\varepsilon$, we have $a+h_1/(\sigma\varepsilon^2)>p_0+1$, and for $\norm{w}\le3$, we have $\sigma\varepsilon\norm{w-\xi}\le r_1$. So the conditions for (\ref{eq:member}) hold, so every site $(w,h)\in Z$ satisfies
\begin{equation}
\label{eq:side}
(w,h)\in\Int H\ \text{ if }h>G(w),\quad\ (w,h)\in\partial H\ \text{ if }h=G(w),\quad\ (w,h)\notin H\ \text{ if }h<G(w).
\end{equation}
Substituting $\tfrac{\sigma}{2}\norm{w-\xi}^2=\tfrac{\sigma}{2}\norm{w}^2-\sigma\langle\xi,w\rangle+\mathcal O(\varepsilon^2)$ and $P(w-\xi)=P(w)+\mathcal O(\varepsilon)$ into (\ref{eq:Gexp}) gives, for $\norm{w}\le3$,
\begin{equation}
\label{eq:exp}
G(w)=p_0+\tfrac{\sigma}{2}\bigl(\norm{w}^2-1\bigr)+\varepsilon\Bigl[\eta(w)+\tfrac{\sigma^2}{6}P(w)\Bigr]+\mathcal O(\varepsilon^2),
\end{equation}
where $\eta(w)\coloneqq\varepsilon^{-1}\bigl(a-p_0+\tfrac{\sigma}{2}-\sigma\langle\xi,w\rangle\bigr)$ is an affine function with $\eta=\mathcal O(1)$ on $\norm{w}\le3$. If $z_k\in\partial H$, then $G(\omega_k)=p_0+\varepsilon\theta_k$ by (\ref{eq:side}), and (\ref{eq:exp}) at $w=\omega_k$ gives
\begin{equation}
\label{eq:expk}
\eta(\omega_k)=\theta_k-\tfrac{\sigma^2}{6}P(\omega_k)+\mathcal O(\varepsilon).
\end{equation}

The proof of Theorem \ref{thm2} uses three lemmas. The following lemma restricts the possible witnesses of $V_{1234}$ to admissible homothets.

\begin{lemma}
\label{lem:four}
Let $H=\sigma^{-1}K_\varepsilon+(\xi,a)$ be a homothet with $z_1,\dots,z_4\in\partial H$ and $b_-,b_+\notin\Int H$. Then
\[
\tfrac13-C\varepsilon\le\sigma\le4+C\varepsilon,\qquad\norm{\xi}\le C\varepsilon,\qquad\bigl\lvert a-p_0+\tfrac{\sigma}{2}\bigr\rvert\le C\varepsilon.
\]
In particular $H$ is admissible.
\end{lemma}

The proof is presented in Appendix \ref{app:four}. It relies on the convexity of $K$ but not its cubic form. 

The next lemma constructs candidate witnesses of the triples $V_{123},..., V_{234}$.

\begin{lemma}
\label{lem:witness}
Let $j\in\{1,\dots,4\}$ and $\sigma_j\in\{\tfrac12,3\}$. Then there is a homothet $W_j=\sigma_j^{-1}K_\varepsilon+(\xi,a)$ with $\norm{\xi}\le C\varepsilon$ and $\lvert a-p_0+\tfrac{\sigma_j}{2}\rvert\le C\varepsilon$ such that $z_k\in\partial W_j$ for the three indices $k\neq j$. In particular $W_j$ is admissible.
\end{lemma}
\begin{proof}
For $k\neq j$, denote the height of the lower graph of $\sigma_j^{-1}K_\varepsilon+(\xi,a)$ above $z_k$ by
\[
F_k(a,\xi;\varepsilon)\coloneqq a+\sigma_j^{-1}f_\varepsilon\bigl(\sigma_j(\omega_k-\xi)\bigr)-p_0-\varepsilon\theta_k.
\]
 Since $f$ is $C^4$ and $f_\varepsilon(y)=\int_0^1(1-t)\,y^T(D^2f)|_{t\varepsilon y}y\,dt$, the $F_k$ extend to $C^2$ functions near $\varepsilon=0$ with $f_0(y)=\tfrac12\norm{y}^2$. At $\varepsilon=0$, the three equations $F_k=0$ are solved by $\xi=0$, $a=p_0-\tfrac{\sigma_j}{2}$, and the Jacobian there has rows $(1,-\sigma_j\omega_k)$, $k\neq j$. It is invertible by (E2), so for a small enough $\varepsilon$ the implicit function theorem gives a solution $(a,\xi)=(p_0-\tfrac{\sigma_j}{2},0)+\mathcal O(\varepsilon)$.
\end{proof}

For each $j$, let $I_j$ be the affine function on $\R^2$ that agrees with $P$ at the three points $\omega_k$, $k\neq j$:
\begin{equation}
\label{eq:I}
I_1\equiv-1,\qquad I_2=w_1+\sqrt3\,w_2,\qquad I_3=-3+4w_1,\qquad I_4=w_1-\sqrt3\,w_2.
\end{equation}
We also have $\bigl(I_j(\omega_*)\bigr)_j=(-1,1,-5,-2)$.

\begin{lemma}
\label{lem:gaps}
Let $j\in\{1,\dots,4\}$, and let $H$ be an admissible homothet with $z_k\in\partial H$ for the three indices $k\neq j$. Then
\begin{enumerate}[(b)]
\item[(a)] $\displaystyle G(\omega_j)=p_0+\varepsilon\theta_j+\varepsilon\,\frac{1-\sigma^2}{\gamma_j}+\mathcal O(\varepsilon^2)$;
\item[(b)] $\displaystyle G((1-\varepsilon)\omega_*)=p_0+\varepsilon\,\beta_j(\sigma)+\mathcal O(\varepsilon^2)$, where $\displaystyle\beta_j(\sigma)\coloneqq\frac56-\sigma+\frac{\sigma^2}{6}+\frac{1-\sigma^2}{6}\,I_j(\omega_*)$.
\end{enumerate}
\end{lemma}
\begin{proof}
The affine function $\eta^*(w)\coloneqq\tfrac12-\tfrac23w_1$ satisfies $\eta^*(\omega_k)=\theta_k-\tfrac16P(\omega_k)$ for all four $k$. So by (\ref{eq:expk}), the affine functions $\eta$ and $\eta^*+\tfrac{1-\sigma^2}{6}I_j$ differ by $\mathcal O(\varepsilon)$ at $\omega_k$ for $k\neq j$, and by (E2)
\begin{equation}
\label{eq:alphaj}
\eta=\eta^*+\tfrac{1-\sigma^2}{6}\,I_j+\mathcal O(\varepsilon)\qquad\text{on }\norm{w}\le3.
\end{equation}
For (a), (\ref{eq:exp}) at $w=\omega_j$ and (\ref{eq:alphaj}) give
\[
G(\omega_j)-p_0-\varepsilon\theta_j=\varepsilon\,\tfrac{1-\sigma^2}{6}\bigl(I_j(\omega_j)-P(\omega_j)\bigr)+\mathcal O(\varepsilon^2),
\]
and (E1) applied to $I_j$ gives $\gamma_j\bigl(I_j(\omega_j)-P(\omega_j)\bigr)=-\sum_{k}\gamma_kP(\omega_k)=6$. For (b), $\norm{(1-\varepsilon)\omega_*}^2-1=-2\varepsilon+\mathcal O(\varepsilon^2)$ and $P((1-\varepsilon)\omega_*)=1+\mathcal O(\varepsilon)$, so (\ref{eq:exp}) at $w=(1-\varepsilon)\omega_*$ gives $G((1-\varepsilon)\omega_*)=p_0+\varepsilon\bigl[-\sigma+\eta(\omega_*)+\tfrac{\sigma^2}{6}\bigr]+\mathcal O(\varepsilon^2)$. By (\ref{eq:alphaj}) and $\eta^*(\omega_*)=\tfrac56$, the bracket is $\beta_j(\sigma)+\mathcal O(\varepsilon)$.
\end{proof}

\begin{proof}[Proof of Theorem \ref{thm2}]
We showed $Z_x=\{z_1,\dots,z_4\}$ in Section \ref{sec:config}.

\emph{Each triple has a witness.} Let $(\sigma_1,\sigma_2,\sigma_3,\sigma_4)\coloneqq(3,\tfrac12,3,\tfrac12)$, and let $W_j$ be the homothet of Lemma \ref{lem:witness}, with lower graph $G_j$. By (\ref{eq:side}), $W_j$ is a witness for $\{1,2,3,4\}\setminus\{j\}$ if the remaining four sites lie below $G_j$. Lemma \ref{lem:gaps} gives the first two rows of Table \ref{tab:heights}, and (\ref{eq:exp}) gives the last two.
\begin{table}[h]
\caption{Height of $G_j$ above each site not on $\partial W_j$}
\label{tab:heights}
\centering
\renewcommand{\arraystretch}{1.3}
\begin{tabular}{l|l|c|c|c|c}
site & height of $G_j$ above the site & $j=1$ & $j=2$ & $j=3$ & $j=4$\\
\hline
$z_j$ & $\varepsilon\,(1-\sigma_j^2)/\gamma_j+\mathcal O(\varepsilon^2)$, with $(1-\sigma_j^2)/\gamma_j=$ & $8/3$ & $3/8$ & $8$ & $3/8$\\
$b_0$ & $\varepsilon\,\bigl(\beta_j(\sigma_j)-\tfrac1{16}\bigr)+\mathcal O(\varepsilon^2)$, with $\beta_j(\sigma_j)-\tfrac1{16}=$ & $29/48$ & $7/16$ & $95/16$ & $1/16$\\
$b_+$ & $\tfrac32\sigma_j-\tfrac12+\mathcal O(\varepsilon)$, with $\tfrac32\sigma_j-\tfrac12=$ & $4$ & $1/4$ & $4$ & $1/4$\\
$b_-$ & $2-\tfrac12\sigma_j+\mathcal O(\varepsilon)$, with $2-\tfrac12\sigma_j=$ & $1/2$ & $7/4$ & $1/2$ & $7/4$
\end{tabular}
\end{table}
All entries are positive, so the four sites lie below $G_j$.

\emph{The quadruple has no witness.} Let $H$ be a homothet with $z_1,\dots,z_4\in\partial H$. If $b_-\in\Int H$ or $b_+\in\Int H$, then $H$ is not a witness. Otherwise $H$ is admissible by Lemma \ref{lem:four}. Since $z_1\in\partial H$, (\ref{eq:side}) gives $G(\omega_1)=p_0+\varepsilon\theta_1$, so Lemma \ref{lem:gaps}(a) with $j=1$ gives $\sigma=1+\mathcal O(\varepsilon)$. As $\beta_1(1)=0$, Lemma \ref{lem:gaps}(b) gives $G((1-\varepsilon)\omega_*)=p_0+\mathcal O(\varepsilon^2)$, which is less than the height $p_0+\tfrac1{16}\varepsilon$ of $b_0$. By (\ref{eq:side}), $b_0\in\Int H$.

All estimates in the proof hold uniformly over all homothets $H$, with margins of order $\varepsilon$, and they continue to hold when the sites in $Z$ are moved by less than $\varepsilon^2$. Hence the conclusion persists when $x$ and the sites are moved by sufficiently small amounts. In particular, we may take $Z$ in general position, and since $Z_x$ is locally constant in $x$, $\chi_x=2$ on an open set of $x$.
\end{proof}

\backmatter

\begin{appendices}

\section{Proofs of Lemmas \ref{planehom} and \ref{homothety}}
\label{app:2d}

\begin{proof}[Proof of Lemma \ref{planehom}]
After rotating, translating, and scaling $K$, we may assume that $x=(0,0,...,-1)$ and $y=(0,0,...,1)$. We take $H$ to be the subset of $\mathbb{R}^d$ where the last coordinate is zero, and $\nu = y$. For any $h\in H$, $v\in \mathbb{R}$ we let $(h,v) \coloneqq h+v\nu$. We will show that $L(x,y)$ intersects each vertical line $l_h \coloneqq \{(h,v): v\in \mathbb{R}\}$ exactly once, and this intersection varies continuously with $h$. Define
\begin{equation}
\begin{split}
u_x \coloneq (-h,-v-1), \quad u_y \coloneq (-h,-v+1), \quad f \coloneq p_K(u_x)-p_K(u_y).
\end{split}
\end{equation}

Let $v\geq 2$, $\lVert h \rVert  \leq v-1$, and $\hat{u}_x$, $\hat{u}_y$ be the unit vectors along $u_x$, $u_y$. Then $\lVert u_x \rVert - \lVert u_y \rVert  = 4v/(\lVert u_x\rVert + \lVert u_y \rVert) \geq 1$ and $\lVert u_x \rVert \leq 2v.$ Since $\arctan$ is $1$-Lipschitz, we have $\lVert \hat{u}_x-\hat{u}_y\rVert \leq \arctan \frac{\lVert h\rVert}{v-1}-\arctan \frac{\lVert h\rVert}{v+1} \leq \frac{3 \lVert h\rVert}{v^2}.$ With homogeneity and Lemma \ref{lemmaLipschitz}, this gives
\begin{equation}
f(h,v) = (\lVert u_x\rVert - \lVert u_y \rVert)p_K(\hat u_y) + \lVert u_x\rVert(p_K(\hat u_x)-p_K(\hat u_y)) \geq p_K(\hat{u}_y) - \frac{6C\lVert h \rVert}{v}.
\end{equation}

For a fixed $h$, we have $\lim_{v \rightarrow \infty} \hat{u}_y = -\nu$, so $\liminf_{v \rightarrow \infty} f(h,v) \geq p_K(-\nu) > 0$. Repeating this argument but exchanging the roles of $u_x$, $u_y$ and letting $v \leq -2$ gives $\limsup_{v \rightarrow -\infty} f(h,v) \leq -p_K(\nu) < 0$. Thus for any fixed $h$, the intermediate value theorem guarantees that $f(h,\cdot)$ has a zero. Next we show that this zero is unique.

Fix $h$ and let $\phi(v) \coloneq p_K(-h,-v)$. Denote its right-derivative by $\phi_+'$, so that $f(h,v) = \int_{v-1}^{v+1}\phi_+'(t) dt$. If $h\neq 0$, then the line $(-h,\cdot)$ misses the origin, so any two points on it are linearly independent. Then by Lemma \ref{lemmabasic}, $\phi$ is strictly convex, and  $\phi'_+$, $f(h,\cdot)$ are strictly increasing. Thus $f(h,\cdot)$ has exactly one zero. If $h=0$, then clearly $f(h,\cdot)$ has exactly one zero. In either case, we call the zero $g(h)$, and $L(x,y) = \{f=0\}$ is the graph of $g$. Next we show that this gives the desired homeomorphism.

Given $h_0$ and $\varepsilon>0$, let $v_{\pm} = g(h_0) \pm \varepsilon$, so that $f(h_0,v_-)<0 < f(h_0,v_+)$. Since $f$ is continuous, we can take $h$ close enough to $h_0$ so that $f(h,v_-) < 0 < f(h,v_+)$. Then $f(h,\cdot)$ must have a zero in $(v_-,v_+)$, which by uniqueness must be $g(h)$. Since $\varepsilon$ was arbitrary, we have shown that $g(h)$ is continuous. Since $(h,g(h)) \mapsto h$ is the restriction of a linear map, it is continuous. Thus $h \mapsto (h,g(h))$ is the desired homeomorphism from $H$ to $L(x,y)$. \end{proof}

\begin{proof}[Proof of Lemma \ref{homothety}]
We adapt the proof of Proposition 14 in \cite{martini2001geometry}. We have $\{x',y',z'\} \coloneqq  h(\{x,y,z\})  \subset h(\partial K')$, and we will assume by contradiction that $\{x',y',z'\} \subset \partial K'$. Let us also assume that the homothety has a center $v$, and assume without loss of generality that $\lambda>1$. If $v \in K'$, then $\lambda>1$ implies $K' \subset h(K')$, so $\partial K' \cap h(\partial K') \subset \{v\}$, contradicting our assumptions on $x',y',$ and $z'$. Otherwise, after translating, rotating, and renaming the points, we may assume that $v=O$ and $x_2,y_2,z_2>0$. By strict convexity, none of the six points may lie in the convex hull of any of other three. Thus no pair of lines from $v$ to $x,y,$ or $z$ may be collinear, since for example, if $vx$ and $vy$ were collinear, then $\{x,y,x',y'\}$ would
lie on a common line. 

After renaming, we may assume that the points are ordered clockwise like $x,y,z$ with $x_2 = z_2$. If $y$ lies above the line $xz$, then $y \in \conv(\{x,y',z\})$, a contradiction. Otherwise, $y'$ lies on the same side of $x'z'$ as $O$, and $y' \in \conv(x',y,z')$, a contradiction.

When $h$ is a translation, we may assume after rotation that it is vertical, $h(A)=A+(0,c)$ with $c>0$. After renaming the points, we can assume that $x_1<y_1<z_1$. If $y$ lies on the same side of $xz$ as $y'$, then $y\in \conv\{x,y',z\}$, and otherwise $y' \in \conv\{x',y,z'\}$. Either one is a contradiction.
\end{proof}

\section{Proof of the Maschke–Pick–Berwald theorem}
\label{app:Pick}

Let $\Omega\subset\mathbb{R}^2$ be a connected open set and $f:\Omega\to\mathbb{R}$ a $C^4$ function with positive definite Hessian. In this section, subscripts denote partial derivatives ($f_i=\partial_i f$, $f_{ij}=\partial_i\partial_j f$), $(f^{ij})$ denotes the inverse of the Hessian $D^2f$, and we observe Einstein notation for summing over repeated indices. In particular,
\begin{equation}\label{eqn_traces}
f^{jk}f_{jk}=2,\qquad f^{jk}f_{ki}=\delta^j_i .
\end{equation}
We set
\[
\varphi \coloneq (\det D^2 f)^{-1/4}>0 .
\]

\begin{lemma}\label{lem_E}
$\kappa=0$ at every point of the graph of $f$ if and only if
\begin{equation}\label{eqn_E}
\varphi f_{ijk}+f_{ij}\varphi_k+f_{jk}\varphi_i+f_{ki}\varphi_j=0
\qquad\text{on }\Omega\text{, for all } i,j,k.
\end{equation}
\end{lemma}
 
\begin{proof}
Fix $w_0\in\Omega$, and in this proof let derivatives of $f$ be evaluated at $w_0$. By Taylor's theorem,
\[
f(w_0+u)=f+f_iu^i+\tfrac12 f_{ij}u^iu^j+\tfrac16 f_{ijk}u^iu^ju^k+\mathcal O(\lVert u\rVert^4).
\]
Let $L$ be an invertible matrix with $L^{T}D^2f(w_0)L=\mathrm{Id}$. Translating $(w_0,f(w_0))$ to the origin, applying the shear $(u,h) \mapsto (u,\,h-f_iu^i)$, and substituting $u=Lw$ gives an affine transformation which brings the graph into the form (\ref{eqn_taylor1}) with
\[
\lVert w\rVert^2=f_{ij}u^iu^j,\qquad c(w)=f_{ijk}u^iu^ju^k .
\]
Now $\kappa=0$ exactly when there is $\alpha\in\mathbb{R}^2$ with
\[
f_{ijk}u^iu^ju^k=3f_{ij}\alpha_k\,u^iu^ju^k\qquad\text{for all }u .
\]
Since $f_{ijk}$ is symmetric, symmetrizing both sides gives
\begin{equation}\label{eqn_alpha}
f_{ijk}=f_{ij}\alpha_k+f_{jk}\alpha_i+f_{ki}\alpha_j .
\end{equation}
Contracting (\ref{eqn_alpha}) with $f^{jk}$ and using (\ref{eqn_traces}) and Jacobi's formula gives
\[
4\alpha_i=f^{jk}f_{ijk}=\partial_i\log\det D^2f=-4\,\varphi_i/\varphi .
\]
Thus $\alpha=-\nabla\varphi/\varphi$, and multiplying (\ref{eqn_alpha}) by $\varphi$ yields (\ref{eqn_E}). Since this held at all possible $w_0$, we are done.
\end{proof}
 
\begin{lemma}\label{lem_affine}
If \emph{(\ref{eqn_E})} holds, then there are constants $\mu,d\in\mathbb{R}$ and $b\in\mathbb{R}^2$ such that
\begin{equation}\label{eqn_phi}
\varphi=\mu f+\langle b,\cdot\rangle+d .
\end{equation}
\end{lemma}
 
\begin{proof}
Differentiate (\ref{eqn_E}) with respect to $w^l$:
\[
\bigl(\varphi_l f_{ijk}+\varphi_k f_{ijl}\bigr)+\varphi f_{ijkl}+f_{jkl}\varphi_i+f_{kil}\varphi_j
+f_{ij}\varphi_{kl}+f_{jk}\varphi_{il}+f_{ki}\varphi_{jl}=0 .
\]
Every term except the last two is symmetric in $k$ and $l$. Thus subtracting the same identity with $k$ and $l$ exchanged leaves
\[
f_{jk}\varphi_{il}+f_{ki}\varphi_{jl}-f_{jl}\varphi_{ik}-f_{li}\varphi_{jk}=0 .
\]
Contracting with $f^{jk}$ and using (\ref{eqn_traces}) gives $2\varphi_{il}+\varphi_{il}-\varphi_{il}-f_{li}f^{jk}\varphi_{jk}=0$, that is,
\begin{equation}\label{eqn_umbilic}
\varphi_{il}=\mu f_{il},\qquad \mu\coloneq\tfrac12 f^{jk}\varphi_{jk}.
\end{equation}
Assume that $\mu$ is $C^1$. Since $\varphi_{ilm}$ and $f_{ilm}$ are symmetric in $l$ and $m$, we get $\mu_m f_{il}=\mu_l f_{im}$. Contracting with $f^{il}$ gives $2\mu_m=\mu_m$, so $\mu$ is constant. A priori $\mu$ is merely continuous, so we instead use the following argument to show that it is constant.

$\nabla f:\Omega\to\mathbb{R}^2$ has invertible Jacobian $(f_{ij})$, so by the inverse function theorem every point of $\Omega$ has a neighborhood $N$ which $\nabla f$ maps diffeomorphically onto an open set containing a disc $D$. Define
$
G\coloneq(\nabla\varphi)\circ (\nabla f)^{-1}:D\to\mathbb{R}^2 ,
$
so that $G(\nabla f(w))=\nabla\varphi(w)$ for $w\in N\cap(\nabla f)^{-1}(D)$. Differentiating component $i$ of this identity with respect to $w^l$ and using (\ref{eqn_umbilic}) gives (here $G_i$ denotes the $i$'th component of $G$, and $\partial^j$ the derivative with respect to the $j$'th coordinate $p_j$ of the variable $p\in D$)
\[
(\partial^j G_i)\,f_{jl}=\varphi_{il}=\mu f_{il}.
\]
Contracting with $f^{lk}$ yields
\[
 \partial^k G_i=\mu\,\delta_i^k.
\]
Hence $\partial^1 G_2 = 0$ and $\partial^2 G_1 = 0$. But then $\mu=\partial^1 G_1$ does not depend on $p_2$, and $\mu=\partial^2 G_2$ does not depend on $p_1$, so $\mu$ is constant on $D$. $\mu$ is locally constant on the connected set $\Omega$, so it is constant. 

By (\ref{eqn_umbilic}), the function $\varphi-\mu f$ has vanishing Hessian, so it is affine.
\end{proof}
 
\begin{proposition}\label{prop_local}
If $\kappa=0$ at every point of the graph of $f$, then the graph is contained in an ellipsoid, an elliptic paraboloid, or a hyperboloid of two sheets.
\end{proposition}
 
\begin{proof}
By Lemmas \ref{lem_E} and \ref{lem_affine}, equations (\ref{eqn_E}), (\ref{eqn_phi}) and (\ref{eqn_umbilic}) hold. Let $S\coloneq\varphi f-\tfrac12\mu f^2$. Using $\varphi_i=\mu f_i+b_i$ we compute
\begin{align*}
S_i&=\varphi f_i+b_i f,\\
S_{ijk}&=\varphi f_{ijk}+\varphi_j f_{ik}+\varphi_k f_{ij}+\varphi_{jk}f_i+b_i f_{jk}
      =f_{jk}\bigl(\mu f_i+b_i-\varphi_i\bigr)=0,
\end{align*}
where the second equality uses (\ref{eqn_E}) and (\ref{eqn_umbilic}). Hence $S(w)$ is a polynomial of degree at most two. Substituting (\ref{eqn_phi}) into the definition of $S$, we find that the graph of $f$ lies in the zero set of
\[
Q(w,h)\coloneq\tfrac12\mu h^2+\bigl(\langle b,w\rangle+d\bigr)h-S(w).
\]
If $Q$ had degree at most one, then $\mu=0$ and $b=0$, so $\varphi=d>0$ and $f=S/d$ would be affine, contradicting $D^2f>0$. So $\{Q=0\}$ is a quadric. The graph has positive Gaussian curvature, and the only quadrics with points of positive curvature are the three listed.
\end{proof}
 
\begin{proof}[Proof of Theorem \ref{thm_pick_berwald}]
Suppose $M$ is an ellipsoid. Since the vanishing of $\kappa$ is affine invariant, we may assume $M$ is the unit sphere, which near each of its points is, up to a rigid motion, the graph of $1-\sqrt{1-\lVert w\rVert^2}=\tfrac12\lVert w\rVert^2+\mathcal O(\lVert w\rVert^4)$. So $\kappa=0$ everywhere.
 
Conversely, suppose $\kappa=0$ on the open set $M_+\subset M$ of points with positive Gaussian curvature. Since $M$ is compact, $M_+$ is nonempty (it contains a point of $M$ farthest from the origin). Let $U$ be one of its connected components. Near any $y\in U$, after an affine transformation, $M$ is the graph of a function $f$ as above. By Proposition \ref{prop_local}, $y$ therefore has a neighborhood in $M$ contained in a quadric of one of the three types above. Such a quadric is determined by any open subset of it, so the quadrics of overlapping neighborhoods coincide, and since $U$ is connected it lies in a single quadric $E$.

Since $M$ and $E$ are closed in $\mathbb{R}^3$, the closure $\overline{U}$ is contained in both. The Gaussian curvatures of $M$ and of $E$ are continuous and agree on $U$, so they agree on $\overline{U}$. As $E$ has positive curvature everywhere, this gives $\overline{U}\subset M_+$. Since $U$ is a connected component of $M_+$, it is closed in $M_+$, and the previous sentence gives $U = \overline{U}$. Since $U$ is open in $M_+$ and $M_+$ is open in $M$, $U$ is open. By the connectedness of $M$, this gives $U=M$. Thus $M$ is a compact surface contained in $E$. $M$ is both open and closed in $E$, so $M=E$ and $E$ must be an ellipsoid.
\end{proof}

\section{Proof of Lemma \ref{lem:four}}
\label{app:four}

We use the notation of Section \ref{sec:3d}. By compactness and strict convexity of $K$, for any $g\in\R^2$ the function $(w,h)\mapsto h-\langle g,w\rangle$ attains its minimum over $K$ at exactly one point. We write this point as $(w_g,h_g)$. For $g=0$ it is the origin.

\begin{lemma}
\label{lem:slope}
There exists $g_1>0$ such that for every $g\in\R^2$ with $\norm{g}\le g_1$,
\begin{equation}
\label{eq:slope}
\nabla f(w_g)=g\qquad\text{and}\qquad\norm{w_g}\le2\norm{g}\le\frac{r_0}{2}.
\end{equation}
\end{lemma}
\begin{proof}
$K$ is compact and the minimizer is unique, so a short argument shows that it depends continuously on $g$. So there is a $g_1$ such that $\norm{w_g}<r_0$ whenever $\norm{g}\le g_1$. After possibly shrinking $g_1$, we may assume that $g_1 \leq r_0/4$. For $\lVert g\rVert \leq g_1$, the definition of $r_0$ gives $h_g=f(w_g)$, so $w_g$ minimizes $f(w)-\langle g,w\rangle$ over $\norm{w}<r_0$. Hence $\nabla f(w_g)=g$. Since $\nabla f(0)=0$ and $D^2f\ge\tfrac12\mathrm{Id}$,
\[
\norm{g}\,\norm{w_g}\ \ge\ \langle\nabla f(w_g)-\nabla f(0),\,w_g\rangle\ \ge\ \tfrac12\norm{w_g}^2,
\]
so $\norm{w_g}\le2\norm{g}\le r_0/2$.
\end{proof}

\begin{lemma}
\label{lem:parab}
Let $H=\sigma^{-1}K_\varepsilon+(\xi,a)$. Let $\tau$ be an affine function of $w$ with $\norm{\nabla\tau}\le g_1/\varepsilon$ such that $H\subset\{h\ge\tau(w)\}$ and the graph of $\tau$ meets $H$ at a point $(w_\zeta,h_\zeta)$. Then $\norm{w_\zeta-\xi}\le2\norm{\nabla\tau}/\sigma$, and with $c_*\coloneqq\min\{\tfrac14,\,r_0/(8\operatorname{diam}K)\}$,
\[
h\ \ge\ \tau(w)+c_*\,\sigma\,\norm{w-w_\zeta}^2\qquad\text{for all }(w,h)\in H.
\]
\end{lemma}
\begin{proof}
Let $g\coloneqq\varepsilon\nabla\tau$, so $\norm{g}\le g_1$. The transformation $A$ given by $Ay \coloneq \sigma^{-1}\Phi_\varepsilon^{-1}y+(\xi,a)$ maps $K$ onto $H$ and multiplies slopes of planes by $1/\varepsilon$. So it sends the plane of slope $g$ supporting $K$ from below to the graph of $\tau$, and $(w_g,h_g)$ to $(w_\zeta,h_\zeta)$. Hence $w_\zeta=\xi+w_g/(\sigma\varepsilon)$, and (\ref{eq:slope}) gives $\norm{w_\zeta-\xi}\le2\norm{g}/(\sigma\varepsilon)=2\norm{\nabla\tau}/\sigma$.

Let $\Psi(w,h)\coloneqq h-h_g-\langle g,w-w_g\rangle$ denote the height of $(w,h)$ above the plane of slope $g$ supporting $K$ at $(w_g,h_g)$. $A$ multiplies heights by $1/(\sigma\varepsilon^2)$ and squared horizontal distances by $1/(\sigma^2\varepsilon^2)$, so the desired inequality on $H$ follows from the inequality $\Psi\ge c_*\norm{w-w_g}^2$ on $K$.  Let $(w,h)\in K$. If $\norm{w}\le r_0$, then $h\ge f(w)$ and $D^2f\ge\tfrac12\mathrm{Id}$ give $\Psi(w,h)\ge\tfrac14\norm{w-w_g}^2$. If $\norm{w}>r_0$, then $\norm{w-w_g}\ge r_0/2$ by (\ref{eq:slope}) and the reverse triangle inequality. Let $q=(w_q,h_q)$ be the point of the segment from $(w_g,h_g)$ to $(w,h)$ with $\norm{w_q-w_g}=r_0/2$. Then $q\in K$ and $\norm{w_q}\le r_0$, so $\Psi(q)\ge r_0^2/16$. As $\Psi$ is affine and vanishes at $(w_g,h_g)$,
\[
\Psi(w,h)=\Psi(q)\,\frac{\norm{w-w_g}}{r_0/2}\ \ge\ \frac{r_0}{8}\norm{w-w_g}\ \ge\ c_*\norm{w-w_g}^2,
\]
where the last inequality used $\norm{w-w_g}\le\operatorname{diam}K$.
\end{proof}

\begin{proof}[Proof of Lemma \ref{lem:four}]
Write $h_k\coloneqq p_0+\varepsilon\theta_k$ for the height of $z_k$.

\emph{Step 1: a supporting plane.} Since $b_-\notin\Int H$ and $H$ is convex, there is a nonconstant affine function $\Psi$ on $\R^3$ with $\Psi\ge0$ on $H$ and $\Psi(b_-)\le0$. Let $\rho\coloneqq(\tfrac15,\tfrac15,\tfrac25,\tfrac15)$, so that $\sum_k\rho_k\omega_k=0$ and $\sum_k\rho_k=1$. The point $\bar z\coloneqq\sum_k\rho_kz_k$ lies in $H$ and directly above $b_-$, so the coefficient of $h$ in $\Psi$ is nonnegative. If it were zero, then $\Psi(\bar z)=\Psi(b_-)\le0$, so $\Psi$ would identically vanish. Hence there is an affine function $\tau$ of $w$ with $H\subset\{h\ge\tau(w)\}$ and $\tau(0)\ge p_0-2$. Adding the largest constant that keeps $H\subset\{h\ge\tau(w)\}$, we may assume that the graph of $\tau$ meets $H$, at a point $(w_\zeta,h_\zeta)$.

Since $z_k\in H$, $\tau(\omega_k)\le h_k\le p_0+1$ for each $k$. Write $\tau(\omega_k)=p_0+1-d_k$ with $d_k\ge0$. As $\tau$ is affine, $\sum_k\rho_k\tau(\omega_k)=\tau(0)\ge p_0-2$, so $\rho_kd_k\le\sum_i\rho_id_i\le3$ and $d_k\le15$ for each $k$. Hence $\tau(\omega_k)\in[p_0-14,\,p_0+1]$. With $\partial_{w_1}\tau=\tfrac12(\tau(\omega_1)-\tau(\omega_3))$ and $\partial_{w_2}\tau=(\tau(\omega_2)-\tau(\omega_4))/\sqrt3$, this gives
\begin{equation}
\label{eq:tau}
0\le h_k-\tau(\omega_k)\le C_0,\qquad\norm{\nabla\tau}\le C_0,\qquad C_0\coloneqq15.
\end{equation}

\emph{Step 2: $\sigma$ is bounded.} Since $\norm{\nabla\tau}\le C_0\le g_1/\varepsilon$, Lemma \ref{lem:parab} applies. At $z_k\in H$ it gives $c_*\sigma\norm{\omega_k-w_\zeta}^2\le h_k-\tau(\omega_k)\le C_0$ for every $k$. Since $\omega_1=-\omega_3$, one of $\omega_1,\omega_3$ is at distance at least $1$ from $w_\zeta$, so $\sigma\le C_0/c_*$. By the first claim of Lemma \ref{lem:parab},
\begin{equation}
\label{eqn_xi_bound}
\norm{\omega_k-\xi}\ \le\ \norm{\omega_k-w_\zeta}+\norm{w_\zeta-\xi}\ \le\ \Bigl(\frac{C_0}{c_*\sigma}\Bigr)^{1/2}+\frac{2C_0}{\sigma}.
\end{equation}
For $\norm{w}\le3$ we have $\norm{w-\xi}\le4+\norm{\omega_k-\xi}$, so, using $\sigma\le C_0/c_*$,
\[
\sigma\varepsilon\norm{w-\xi}\ \le\ \varepsilon\bigl(4\sigma+\sqrt{C_0\sigma/c_*}+2C_0\bigr)\ \le\ C\varepsilon\ \le\ r_1.
\]
So (\ref{eq:Gexp}) and (\ref{eq:member}) apply on $\norm{w}\le3$. (\ref{eqn_xi_bound}) gives $\lVert \xi \rVert \leq C(1+1/\sigma),$ and $(\xi,a)$ lies in $H$, so $a\ge\tau(\xi)\ge\tau(0)-C_0\norm{\xi}\ge-C(1+1/\sigma)$. Therefore for small enough $\varepsilon$, the upper limit $a+h_1/(\sigma\varepsilon^2)$ in (\ref{eq:member}) exceeds $p_0+1$. Each $z_k$ lies on $\partial H$ at height at most $p_0+1$, so by (\ref{eq:member}) this gives $G(\omega_k)=h_k$.

\emph{Step 3: improved $\sigma$ bound.} By (\ref{eq:Gexp}), $D^2G=\sigma(\mathrm{Id}+\mathcal O(\varepsilon))$ on $\norm{w}\le3$, so $G-\tfrac{\sigma}{2}\norm{w}^2$ differs from an affine function by $\mathcal O(\varepsilon)$ there. That affine function takes the values $p_0-\sigma/2 + \mathcal O(\varepsilon)$ at the $\omega_k$, so by (E2), for $\lVert w\rVert \leq 3$,
\begin{equation}
\label{eq:order0}
G(w)=p_0+\tfrac{\sigma}{2}\bigl(\norm{w}^2-1\bigr)+\mathcal O(\varepsilon).
\end{equation}
At $w=2\omega_*$ this is $p_0+\tfrac32\sigma+\mathcal O(\varepsilon)$, and at $w=0$ it is $p_0-\tfrac{\sigma}{2}+\mathcal O(\varepsilon)$. By (\ref{eq:member}), $b_+\notin\Int H$ gives $\tfrac32\sigma\ge\tfrac12-C\varepsilon$, and $b_-\notin\Int H$ gives $\tfrac{\sigma}{2}\le2+C\varepsilon$.

\emph{Step 4: $\xi$ and $a$ bounds.} Now $\tfrac14\le\sigma\le5$ and $\norm{\xi}\le C$, so on $\norm{w}\le3$, (\ref{eq:Gexp}) gives $G(w)=a+\tfrac{\sigma}{2}\norm{w-\xi}^2+\mathcal O(\varepsilon)$. Subtracting (\ref{eq:order0}) shows that the affine function $a+\tfrac{\sigma}{2}\norm{\xi}^2-p_0+\tfrac{\sigma}{2}-\sigma\langle\xi,w\rangle$ is $\mathcal O(\varepsilon)$ on $\norm{w}\le3$. Its gradient is $-\sigma\xi$ and $\sigma\ge\tfrac14$, so $\xi=\mathcal O(\varepsilon)$. Plugging this bound on $\xi$ back into the affine function gives $a-p_0+\tfrac{\sigma}{2}=\mathcal O(\varepsilon)$.
\end{proof}

\end{appendices}

\backmatter

\bmhead{Acknowledgements}
The authors thank Henry Bosch for helpful discussions and suggestions.

\section*{Declarations}

\bmhead{Funding}
This research was supported in part by the Air Force Office of Scientific Research under grants FA9550-21-1-0381 and FA9550-24-1-0237, and by the Office of Advanced Scientific Computing Research (ASCR) within the Department of Energy Office of Science under award number DE-SC0023163.

\bmhead{Competing interests}
The authors declare no competing interests.

\bmhead{Data availability}
Data sharing not applicable to this article as no datasets were generated or analyzed during the current study.

\bmhead{Code availability}
The code used to generate the figures is available from the corresponding author on request.

\bibliography{Union_Area_combined}


\begin{thebibliography}{19}
\ifx \bisbn   \undefined \def \bisbn  #1{ISBN #1}\fi
\ifx \binits  \undefined \def \binits#1{#1}\fi
\ifx \bauthor  \undefined \def \bauthor#1{#1}\fi
\ifx \batitle  \undefined \def \batitle#1{#1}\fi
\ifx \bjtitle  \undefined \def \bjtitle#1{#1}\fi
\ifx \bvolume  \undefined \def \bvolume#1{\textbf{#1}}\fi
\ifx \byear  \undefined \def \byear#1{#1}\fi
\ifx \bissue  \undefined \def \bissue#1{#1}\fi
\ifx \bfpage  \undefined \def \bfpage#1{#1}\fi
\ifx \blpage  \undefined \def \blpage #1{#1}\fi
\ifx \burl  \undefined \def \burl#1{\textsf{#1}}\fi
\ifx \doiurl  \undefined \def \doiurl#1{\url{https://doi.org/#1}}\fi
\ifx \betal  \undefined \def \betal{\textit{et al.}}\fi
\ifx \binstitute  \undefined \def \binstitute#1{#1}\fi
\ifx \binstitutionaled  \undefined \def \binstitutionaled#1{#1}\fi
\ifx \bctitle  \undefined \def \bctitle#1{#1}\fi
\ifx \beditor  \undefined \def \beditor#1{#1}\fi
\ifx \bpublisher  \undefined \def \bpublisher#1{#1}\fi
\ifx \bbtitle  \undefined \def \bbtitle#1{#1}\fi
\ifx \bedition  \undefined \def \bedition#1{#1}\fi
\ifx \bseriesno  \undefined \def \bseriesno#1{#1}\fi
\ifx \blocation  \undefined \def \blocation#1{#1}\fi
\ifx \bsertitle  \undefined \def \bsertitle#1{#1}\fi
\ifx \bsnm \undefined \def \bsnm#1{#1}\fi
\ifx \bsuffix \undefined \def \bsuffix#1{#1}\fi
\ifx \bparticle \undefined \def \bparticle#1{#1}\fi
\ifx \barticle \undefined \def \barticle#1{#1}\fi
\bibcommenthead
\ifx \bconfdate \undefined \def \bconfdate #1{#1}\fi
\ifx \botherref \undefined \def \botherref #1{#1}\fi
\ifx \url \undefined \def \url#1{\textsf{#1}}\fi
\ifx \bchapter \undefined \def \bchapter#1{#1}\fi
\ifx \bbook \undefined \def \bbook#1{#1}\fi
\ifx \bcomment \undefined \def \bcomment#1{#1}\fi
\ifx \oauthor \undefined \def \oauthor#1{#1}\fi
\ifx \citeauthoryear \undefined \def \citeauthoryear#1{#1}\fi
\ifx \endbibitem  \undefined \def \endbibitem {}\fi
\ifx \bconflocation  \undefined \def \bconflocation#1{#1}\fi
\ifx \arxivurl  \undefined \def \arxivurl#1{\textsf{#1}}\fi
\csname PreBibitemsHook\endcsname

%%% 1
\bibitem[\protect\citeauthoryear{Naiman and Wynn}{1992}]{naiman1992inclusion}
\begin{barticle}
\bauthor{\bsnm{Naiman}, \binits{D.Q.}},
\bauthor{\bsnm{Wynn}, \binits{H.P.}}:
\batitle{Inclusion-exclusion-{B}onferroni identities and inequalities for discrete tube-like problems via {E}uler characteristics}.
\bjtitle{Ann. Statist.}
\bvolume{20}(\bissue{1}),
\bfpage{43}--\blpage{76}
(\byear{1992})
\doiurl{10.1214/aos/1176348512}
\end{barticle}
\endbibitem

%%% 2
\bibitem[\protect\citeauthoryear{Edelsbrunner}{1995}]{edelsbrunner1995union}
\begin{barticle}
\bauthor{\bsnm{Edelsbrunner}, \binits{H.}}:
\batitle{The union of balls and its dual shape}.
\bjtitle{Discrete Comput. Geom.}
\bvolume{13}(\bissue{3--4}),
\bfpage{415}--\blpage{440}
(\byear{1995})
\doiurl{10.1007/BF02574053}
\end{barticle}
\endbibitem

%%% 3
\bibitem[\protect\citeauthoryear{Attali and Edelsbrunner}{2007}]{attali2007inclusion}
\begin{barticle}
\bauthor{\bsnm{Attali}, \binits{D.}},
\bauthor{\bsnm{Edelsbrunner}, \binits{H.}}:
\batitle{Inclusion-exclusion formulas from independent complexes}.
\bjtitle{Discrete Comput. Geom.}
\bvolume{37}(\bissue{1}),
\bfpage{59}--\blpage{77}
(\byear{2007})
\doiurl{10.1007/s00454-006-1274-7}
\end{barticle}
\endbibitem

%%% 4
\bibitem[\protect\citeauthoryear{Aurenhammer and Klein}{2000}]{aurenhammer2000voronoi}
\begin{bchapter}
\bauthor{\bsnm{Aurenhammer}, \binits{F.}},
\bauthor{\bsnm{Klein}, \binits{R.}}:
\bctitle{Voronoi diagrams}.
In: \beditor{\bsnm{Sack}, \binits{J.-R.}},
\beditor{\bsnm{Urrutia}, \binits{J.}} (eds.)
\bbtitle{Handbook of Computational Geometry},
pp. \bfpage{201}--\blpage{290}.
\bpublisher{North-Holland},
\blocation{Amsterdam}
(\byear{2000}).
\doiurl{10.1016/B978-044482537-7/50006-1}
\end{bchapter}
\endbibitem

%%% 5
\bibitem[\protect\citeauthoryear{Ma}{2000}]{ma2000bisectors}
\begin{botherref}
\oauthor{\bsnm{Ma}, \binits{L.}}:
Bisectors and {V}oronoi diagrams for convex distance functions.
PhD thesis,
FernUniversit{\"a}t Hagen
(2000)
\end{botherref}
\endbibitem

%%% 6
\bibitem[\protect\citeauthoryear{Chew and Drysdale}{1985}]{chew1985voronoi}
\begin{bchapter}
\bauthor{\bsnm{Chew}, \binits{L.P.}},
\bauthor{\bsnm{Drysdale}, \binits{R.L.S.} \bsuffix{III}}:
\bctitle{Voronoi diagrams based on convex distance functions}.
In: \bbtitle{Proceedings of the First Annual Symposium on Computational Geometry},
pp. \bfpage{235}--\blpage{244}.
\bpublisher{ACM},
\blocation{New York}
(\byear{1985}).
\doiurl{10.1145/323233.323264}
\end{bchapter}
\endbibitem

%%% 7
\bibitem[\protect\citeauthoryear{Corbal{\'a}n et~al.}{1996}]{corbalan1996geometry}
\begin{barticle}
\bauthor{\bsnm{Corbal{\'a}n}, \binits{A.G.}},
\bauthor{\bsnm{Maz{\'o}n}, \binits{M.}},
\bauthor{\bsnm{Recio}, \binits{T.}}:
\batitle{Geometry of bisectors for strictly convex distances}.
\bjtitle{Internat. J. Comput. Geom. Appl.}
\bvolume{6}(\bissue{1}),
\bfpage{45}--\blpage{58}
(\byear{1996})
\doiurl{10.1142/S0218195996000046}
\end{barticle}
\endbibitem

%%% 8
\bibitem[\protect\citeauthoryear{Aurenhammer et~al.}{2013}]{aurenhammer2013voronoi}
\begin{bbook}
\bauthor{\bsnm{Aurenhammer}, \binits{F.}},
\bauthor{\bsnm{Klein}, \binits{R.}},
\bauthor{\bsnm{Lee}, \binits{D.-T.}}:
\bbtitle{Voronoi Diagrams and {D}elaunay Triangulations}.
\bpublisher{World Scientific},
\blocation{Hackensack, NJ}
(\byear{2013}).
\doiurl{10.1142/8685}
\end{bbook}
\endbibitem

%%% 9
\bibitem[\protect\citeauthoryear{Edelsbrunner and Ramos}{1997}]{edelsbrunner1997inclusion}
\begin{barticle}
\bauthor{\bsnm{Edelsbrunner}, \binits{H.}},
\bauthor{\bsnm{Ramos}, \binits{E.A.}}:
\batitle{Inclusion-exclusion complexes for pseudodisk collections}.
\bjtitle{Discrete Comput. Geom.}
\bvolume{17}(\bissue{3}),
\bfpage{287}--\blpage{306}
(\byear{1997})
\doiurl{10.1007/PL00009295}
\end{barticle}
\endbibitem

%%% 10
\bibitem[\protect\citeauthoryear{Borsuk}{1948}]{borsuk1948imbedding}
\begin{barticle}
\bauthor{\bsnm{Borsuk}, \binits{K.}}:
\batitle{On the imbedding of systems of compacta in simplicial complexes}.
\bjtitle{Fund. Math.}
\bvolume{35},
\bfpage{217}--\blpage{234}
(\byear{1948})
\doiurl{10.4064/fm-35-1-217-234}
\end{barticle}
\endbibitem

%%% 11
\bibitem[\protect\citeauthoryear{Goodey and Woodcock}{1981}]{goodey1981intersections}
\begin{bchapter}
\bauthor{\bsnm{Goodey}, \binits{P.R.}},
\bauthor{\bsnm{Woodcock}, \binits{M.M.}}:
\bctitle{Intersections of convex bodies with their translates}.
In: \beditor{\bsnm{Davis}, \binits{C.}},
\beditor{\bsnm{Gr{\"u}nbaum}, \binits{B.}},
\beditor{\bsnm{Sherk}, \binits{F.A.}} (eds.)
\bbtitle{The Geometric Vein},
pp. \bfpage{289}--\blpage{296}.
\bpublisher{Springer},
\blocation{New York}
(\byear{1981}).
\doiurl{10.1007/978-1-4612-5648-9_21}
\end{bchapter}
\endbibitem

%%% 12
\bibitem[\protect\citeauthoryear{Goodey}{1983}]{Goodey_1983}
\begin{barticle}
\bauthor{\bsnm{Goodey}, \binits{P.R.}}:
\batitle{Homothetic ellipsoids}.
\bjtitle{Math. Proc. Cambridge Philos. Soc.}
\bvolume{93}(\bissue{1}),
\bfpage{25}--\blpage{34}
(\byear{1983})
\doiurl{10.1017/S0305004100060291}
\end{barticle}
\endbibitem

%%% 13
\bibitem[\protect\citeauthoryear{Santos}{1996}]{santos1996delaunay}
\begin{barticle}
\bauthor{\bsnm{Santos}, \binits{F.}}:
\batitle{On {D}elaunay oriented matroids for convex distance functions}.
\bjtitle{Discrete Comput. Geom.}
\bvolume{16}(\bissue{2}),
\bfpage{197}--\blpage{210}
(\byear{1996})
\doiurl{10.1007/BF02716807}
\end{barticle}
\endbibitem

%%% 14
\bibitem[\protect\citeauthoryear{Icking et~al.}{1999}]{icking1999bisectors}
\begin{bchapter}
\bauthor{\bsnm{Icking}, \binits{C.}},
\bauthor{\bsnm{Klein}, \binits{R.}},
\bauthor{\bsnm{L{\^e}}, \binits{N.-M.}},
\bauthor{\bsnm{Ma}, \binits{L.}},
\bauthor{\bsnm{Santos}, \binits{F.}}:
\bctitle{On bisectors for convex distance functions in 3-space}.
In: \bbtitle{Proc. 11th Canadian Conference on Computational Geometry (CCCG'99)}
(\byear{1999})
\end{bchapter}
\endbibitem

%%% 15
\bibitem[\protect\citeauthoryear{Icking et~al.}{1995}]{icking1995convex}
\begin{barticle}
\bauthor{\bsnm{Icking}, \binits{C.}},
\bauthor{\bsnm{Klein}, \binits{R.}},
\bauthor{\bsnm{L{\^e}}, \binits{N.-M.}},
\bauthor{\bsnm{Ma}, \binits{L.}}:
\batitle{Convex distance functions in 3-space are different}.
\bjtitle{Fund. Inform.}
\bvolume{22}(\bissue{4}),
\bfpage{331}--\blpage{352}
(\byear{1995})
\doiurl{10.3233/FI-1995-2242}
\end{barticle}
\endbibitem

%%% 16
\bibitem[\protect\citeauthoryear{Nomizu and Sasaki}{1994}]{NomizuSasaki1994}
\begin{bbook}
\bauthor{\bsnm{Nomizu}, \binits{K.}},
\bauthor{\bsnm{Sasaki}, \binits{T.}}:
\bbtitle{Affine Differential Geometry: Geometry of Affine Immersions}.
\bsertitle{Cambridge Tracts in Mathematics},
vol. \bseriesno{111}.
\bpublisher{Cambridge University Press},
\blocation{Cambridge}
(\byear{1994})
\end{bbook}
\endbibitem

%%% 17
\bibitem[\protect\citeauthoryear{Martini et~al.}{2001}]{martini2001geometry}
\begin{barticle}
\bauthor{\bsnm{Martini}, \binits{H.}},
\bauthor{\bsnm{Swanepoel}, \binits{K.J.}},
\bauthor{\bsnm{Wei{\ss}}, \binits{G.}}:
\batitle{The geometry of {M}inkowski spaces---a survey. {P}art {I}}.
\bjtitle{Expo. Math.}
\bvolume{19}(\bissue{2}),
\bfpage{97}--\blpage{142}
(\byear{2001})
\doiurl{10.1016/S0723-0869(01)80025-6}
\end{barticle}
\endbibitem

%%% 18
\bibitem[\protect\citeauthoryear{Martini and Swanepoel}{2004}]{martini2004geometry}
\begin{barticle}
\bauthor{\bsnm{Martini}, \binits{H.}},
\bauthor{\bsnm{Swanepoel}, \binits{K.J.}}:
\batitle{The geometry of {M}inkowski spaces---a survey. {P}art {II}}.
\bjtitle{Expo. Math.}
\bvolume{22}(\bissue{2}),
\bfpage{93}--\blpage{144}
(\byear{2004})
\doiurl{10.1016/S0723-0869(04)80009-4}
\end{barticle}
\endbibitem

%%% 19
\bibitem[\protect\citeauthoryear{Blaschke}{1923}]{blaschke1923affine}
\begin{bbook}
\bauthor{\bsnm{Blaschke}, \binits{W.}}:
\bbtitle{Vorlesungen {\"U}ber {D}ifferentialgeometrie und Geometrische {G}rundlagen Von {E}insteins {R}elativit{\"a}tstheorie. {II}: {A}ffine {D}ifferentialgeometrie}.
\bsertitle{Die Grundlehren der mathematischen Wissenschaften},
vol. \bseriesno{7}.
\bpublisher{Springer},
\blocation{Berlin}
(\byear{1923}).
\bcomment{Bearbeitet von K. Reidemeister}
\end{bbook}
\endbibitem

\end{thebibliography}

\end{document}